# Moiré metrology of energy landscapes in van der Waals heterostructures

Dorri Halbertal[1*], Nathan R. Finney[2], Sai S. Sunku[1], Alexander Kerelsky[1], Carmen Rubio-Verdú[1], Sara Shabani[1], Lede Xian[3,4], Stephen Carr[5,6], Shaowen Chen[1,7], Charles Zhang[1,8], Lei Wang[1], Derick Gonzalez-Acevedo[1,7], Alexander S. McLeod[1], Daniel Rhodes[1,9], Kenji Watanabe[10], Takashi Taniguchi[10], Efthimios Kaxiras[11], Cory R. Dean[1], James C. Hone[2], Abhay N. Pasupathy[1], Dante M. Kennes[3,12], Angel Rubio[3,13], D. N. Basov[1]

## Abstract

The emerging field of twistronics, which harnesses the twist angle between two-dimensional materials, represents a promising route for the design of quantum materials, as the twist-angle-induced superlattices offer means to control topology and strong correlations. At the small twist limit, and particularly under strain, as atomic relaxation prevails, the emergent moiré superlattice encodes elusive insights into the local interlayer interaction. Here we introduce moiré metrology as a combined experiment-theory framework to probe the stacking energy landscape of bilayer structures at the 0.1 meV/atom scale, outperforming the gold-standard of quantum chemistry. Through studying the shapes of moiré domains with numerous nano-imaging techniques, and correlating with multi-scale modelling, we assess and refine first-principle models for the interlayer interaction. We document the prowess of moiré metrology for three representative twisted systems: bilayer graphene, double bilayer graphene and H-stacked $MoSe_2/WSe_2$. Moiré metrology establishes sought after experimental benchmarks for interlayer interaction, thus enabling accurate modelling of twisted multilayers.

## Introduction

Twisted van der Waals structures, such as twisted bilayer graphene[1–16] (TBG), twisted double bilayer graphene[17–19] (TDBG) and twisted transition-metal-dichalcogenides[20–27] are in the vanguard of quantum materials research[28–30]. The twist between the layers leads to large-scale periodic perturbations of stacking configurations, called a moiré superlattice. Because atomic layers in van der Waals (vdW) materials are not rigid but instead behave as deformable membranes, moiré suprelattices acquire additional attributes. As two atomic layers with a small relative twist angle come in contact, the atomic positions relax to minimize the total energy. Through the relaxation process domains of lowest energy configurations form and become separated by domain walls of transitionary configurations[31–33] (Fig. 1a). The generalized stacking fault energy function (GSFE), which provides the energetic variations across different stacking configurations, is the fundamental property that describes relaxed vdW interfaces[31,34]. The GSFE is commonly calculated using density functional theory (DFT)[31,34]. Experimental techniques[35] to probe the GSFE are currently restricted to the stable lowest energy configuration, and are very limited in energy resolution compared to the variability among theoretical descriptions.

[1]Department of Physics, Columbia University, New York, NY, USA.
[2]Department of Mechanical Engineering, Columbia University, New York, NY, USA.
[3]Max Planck Institute for the Structure and Dynamics of Matter and Center Free-Electron Laser Science, Luruper Chaussee 149, 22761 Hamburg, Germany.
[4]Present address: Songshan Lake Materials Laboratory, Dongguan, Guangdong 523808, China.
[5]Department of Physics, Harvard University, Cambridge, Massachusetts 02138, USA.
[6]Present address: Brown University, Providence, RI 02912, USA.
[7]Present address: Department of Physics, Harvard University, Cambridge, MA 02138, USA.
[8]Present address: Department of Physics, University of California at Santa Barbara, Santa Barbara, CA 93106, USA.
[9]Present address: Department of Materials Science and Engineering, University of Wisconsin-Madison, WI 53706, USA.
[10]National Institute for Material Science, Tsukuba, Japan
[11]John A. Paulson School of Engineering and Applied Sciences, Harvard University, Cambridge, Massachusetts 02138, USA.
[12]Institut fur Theorie der Statistischen Physik, RWTH Aachen University, 52056 Aachen, Germany.
[13]Center for Computational Quantum Physics, Flatiron Institute, New York, NY 10010 USA.
*Correspondence and requests for materials should be addressed to D.H. (dh2917@columbia.edu).

Here we show that the generalized stacking fault energy function (GSFE) is encoded in fine details of the relaxed moiré super-lattice patterns at the low twist-angle limit. In particular, the shape of domains and domain walls networks, as well as domain wall width, abide by transitionary configurations beyond the lowest-energy stackings of the domains. More specifically, we distinguish between single and double domain walls (SDW and DDW). SDWs separate two distinct stacking configurations of a moiré superlattice (for instance, ABCA [MM'] and ABAB [MX'] in the TDBG [for twisted H-stacked $MoSe_2/WSe_2$, or T-H-$MoSe_2/WSe_2$ for short] example of Fig. 1a). DDWs, formed from the collapse of two SDWs, separate identical phases (ABAB for TDBG and MX' for T-H-$MoSe_2/WSe_2$ in Fig. 1a). The formation and nature of DDWs result from attraction of SDWs as they are brought together (for instance, due to external or relaxation induced strain), and is proven here to provide a reliable read-out of the underlying energetics. In cases of inequivalent two lowest energy configurations (as in Fig. 1), the SDW develops a finite curvature $\kappa$, allowing one to extract the domains energy imbalance with an accuracy outperforming the $\sim 3\, meV/atom$ of the gold standard of quantum chemistry[36,37].

## Results

Moiré metrology, presented here, correlates measurable spatial patterns of the relaxed moiré superlattice (such as shapes of domains, SDWs and formation of DDWs) with modelling based on the GSFE. To do so we developed a continuous two dimensional relaxation simulation. The model searches for local inter-layer displacement fields that minimize the total energy of the multilayer, as a sum of elastic and stacking energy terms (see SI sections S1-2 for more details, also see Ref. 33 for an alternative approach). The equations are solved in real space and thus capture subtle experimental details that remained underexplored. Fig. 1b-g is a tour-de-force of moiré metrology combining experimental imaging of different systems, techniques and length-scales (Fig. 1b-d), and their respective modelling (Fig. 1e-g). Fig. 1b-d were acquired with modern scanning probe microscopy (SPM) techniques: scanning tunneling microscopy (constant current mode) and spectroscopy[17] (STM and STS respectively) and mid-infrared range (mid-IR) scanning nearfield optical microscopy[11] (SNOM). These techniques resolve stacking configurations based on local topographic and electronic (STM and STS), as well as mid-IR optical conductivity (mid-IR SNOM) contrasts. In low strain TDBG, the model (Fig. 1e) captures the fine curving of SDWs (Fig. 1b). In cases of higher strain (Fig. 1c and modelling in Fig. 1f) we observe the formation of one dimensional DDW structures (inset of Fig. 1f highlights an example). Similarly, DDW formations and SDW curving were observed (Fig. 1d) and modelled (Fig. 1g) in T-H-$MoSe_2/WSe_2$, with excellent agreement across different length-scales of the image (see SI S3 for additional analysis). Next we will illustrate in detail how moiré super-lattices reveal the energy landscape information using TBG and TDBG as prototypical examples.

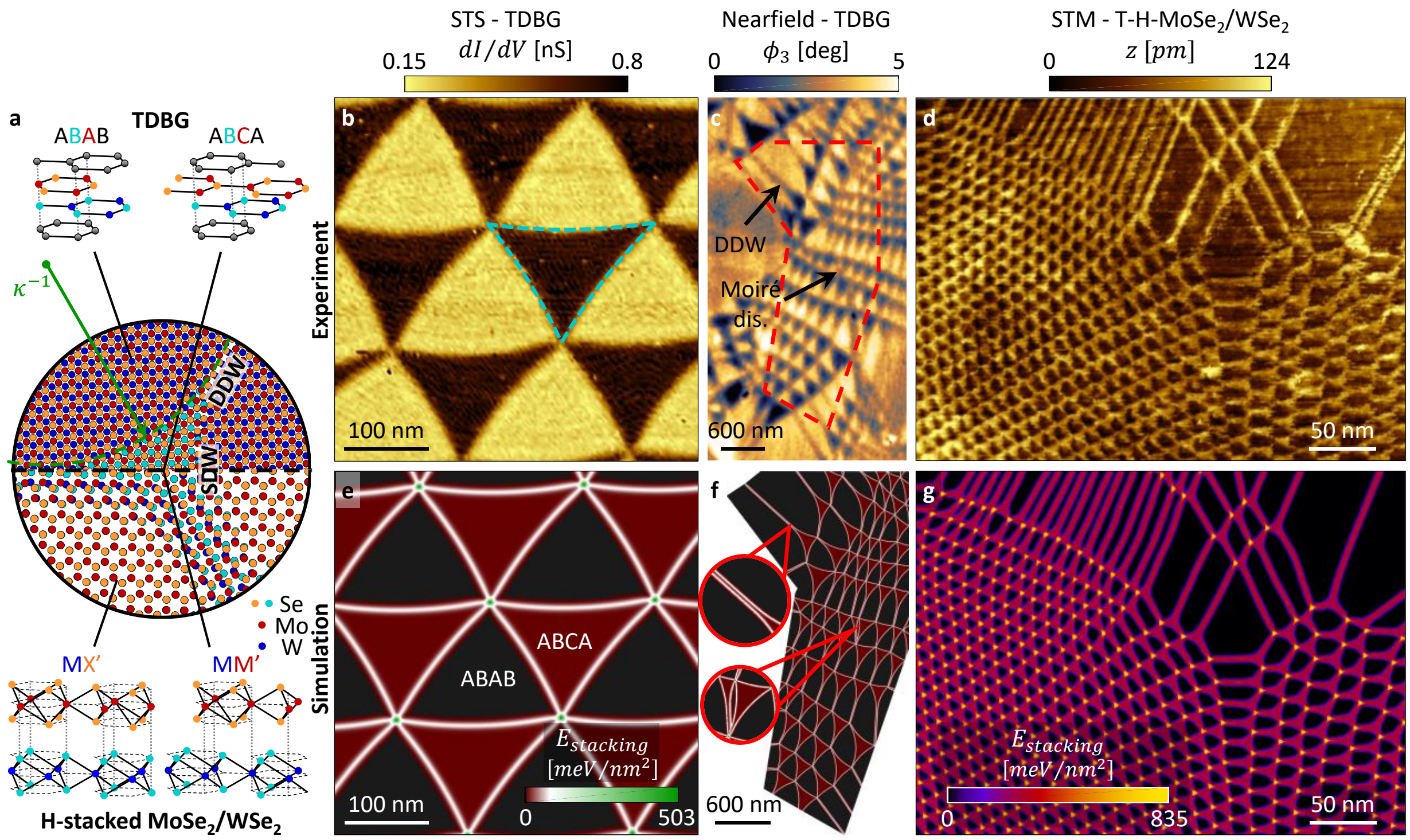

**Figure 1|Physics of atomic layers stacking probed by moiré metrology in vdW twisted bilayers – experiment and theory. a,** Illustration of domain formations in a relaxed twisted bilayer structure. Center: atomic positioning after relaxation (see SI S1 for more details). Atoms are colored to highlight stacking configurations. The energy imbalance leads to curved single domain walls (SDWs), with radius of curvature indicated by $\kappa^{-1}$, and in some cases with formation of double domain walls (DDWs). Two systems with energy imbalance are considered: TDBG (top) and T-H-$MoSe_2$/$WSe_2$ (bottom). **b,** STS map of TDBG with $\theta = 0.07°$, revealing rhombohedral (ABCA – dark) and Bernal (ABAB – bright) domains with minimal external strain. The rhombohedral phase bends inward (dashed turquoise line) revealing an energy imbalance between the two phases as discussed in the text. **c,** Mid-IR (940 $\mathrm{cm}^{-1}$) nearfield phase imaging of TDBG resolves ABCA (dark) and ABAB (bright) phases and DDW formations. **d,** STM map of T-H-$MoSe_2$/$WSe_2$ resolving MM' (bright) and MX' (dark) stacking configurations as well as DDW formation in various strain conditions. **e-g,** Stacking energy density from full relaxation calculations of the experimental cases of **b-d** respectively (see methods, SI sections S1-2 and text for more details). The color-map is shared for **e-f**. Magnified regions in **f** (and arrows in **c**) highlight a DDW formation and a moiré dislocation (see discussion in SI S3). Calculated region of **f** is marked by dashed shape in **c**.

## Moiré metrology of twisted bilayer graphene

To study the energy landscape of TBG, we focus on the interplay between SDW and DDW formations. Fig. 2a presents a non-local nano-photocurrent map of TBG in the minimal twist limit <$0.1^0$. Bright spots in the photo-current map highlight the AA sites (indicating higher absorption – see methods and SI S4). The AA sites are connected by domain walls separating AB and BA domains. The resultant moiré super-lattice is clearly affected by strain, inferred from the distorted triangular pattern, especially near the edges of the stack. There, we observe the merging of two SDWs into a single DDW (selected locations are marked in Fig. 2a). We successfully account for the observed network within a model addressing a competition between SDWs and DDWs. To grasp the essential physics, we first assume a characteristic energy of forming a segment of DDW and SDW. We define a dimensionless domain-wall formation ratio as the ratio of DDW and SDW line energies, $\bar{\beta} = \gamma_{DDW}/\gamma_{SDW}$. In addition to $\bar{\beta}$, the model input includes the AA sites of the moiré pattern as the fixed vertices of the triangles forming the network. We explore the SDW vs. DDW structures that emerge for a given value of the single tuning parameter $\bar{\beta}$. The case of $\bar{\beta} = 2$ implies there is no benefit in forming a DDW, and the optimal structure would simply be straight SDWs connecting the AA sites. For $\bar{\beta} < 2$ the two SDWs attract each other favoring the emergence of DDW segments (see SI S4 and Supplementary Video 1 for details). Our modeling captures the overall shape of the experimental map for $\bar{\beta} = 1.90$ (Fig. 2a) The agreement is remarkable considering the minimal modeling we employ. We conclude that in order for a TBG model to reproduce the experimental picture, two SDWs have to sufficiently attract one another as quantified by the fitted $\bar{\beta}$. In that sense, as we show more rigorously below, moiré metrology puts constraints the GSFE.

To quantify how the observed moiré networks constrain the stacking energy landscape, we span all realistic GSFE's satisfying the symmetry of TBG over a 2D unit-less parameter space $(\zeta, \tau)$ (as illustrated in Fig. 2b and discussed at SI S4), such that each point on the $(\zeta, \tau)$ plane represents one GSFE candidate. We solve a set of 1D relaxation problems describing the profiles of SDWs and DDWs (see SI S1 for more details on relaxation codes), and extract the domain wall formation ratio $\bar{\beta}$. This allows us to define a band in $(\zeta, \tau)$ plane of GSFE's that comply with the experimental $\bar{\beta} = 1.90$ (see SI Fig. S3g). Fig. 2b compares one GSFE moiré constrained candidate with the $\bar{\beta} = 1.90$ band (magenta) with the well-accepted choice of GSFE of Ref. 31 (blue), which notably falls outside of the band with $\bar{\beta} = 1.98$. The moiré metrology analysis indicates that SDWs implied by GSFE in Ref. 31 insufficiently attract one another (blue curve in Fig. 2c) to account for the observed network, as indeed revealed in Fig. 2d. In contrast, the moiré-constrained candidate (magenta in Fig. 2b) with a flatter saddle point promotes stronger SDWs attraction across a broad range of domain wall orientations (magenta curve in Fig. 2c; see SI sections S1 and S5 for additional details), and yields excellent agreement with the data (Fig. 2e). Regardless of the good agreement, the relatively flat saddle point comes as a surprise, and may in fact correct for an unknown effect unrelated to interlayer energy.

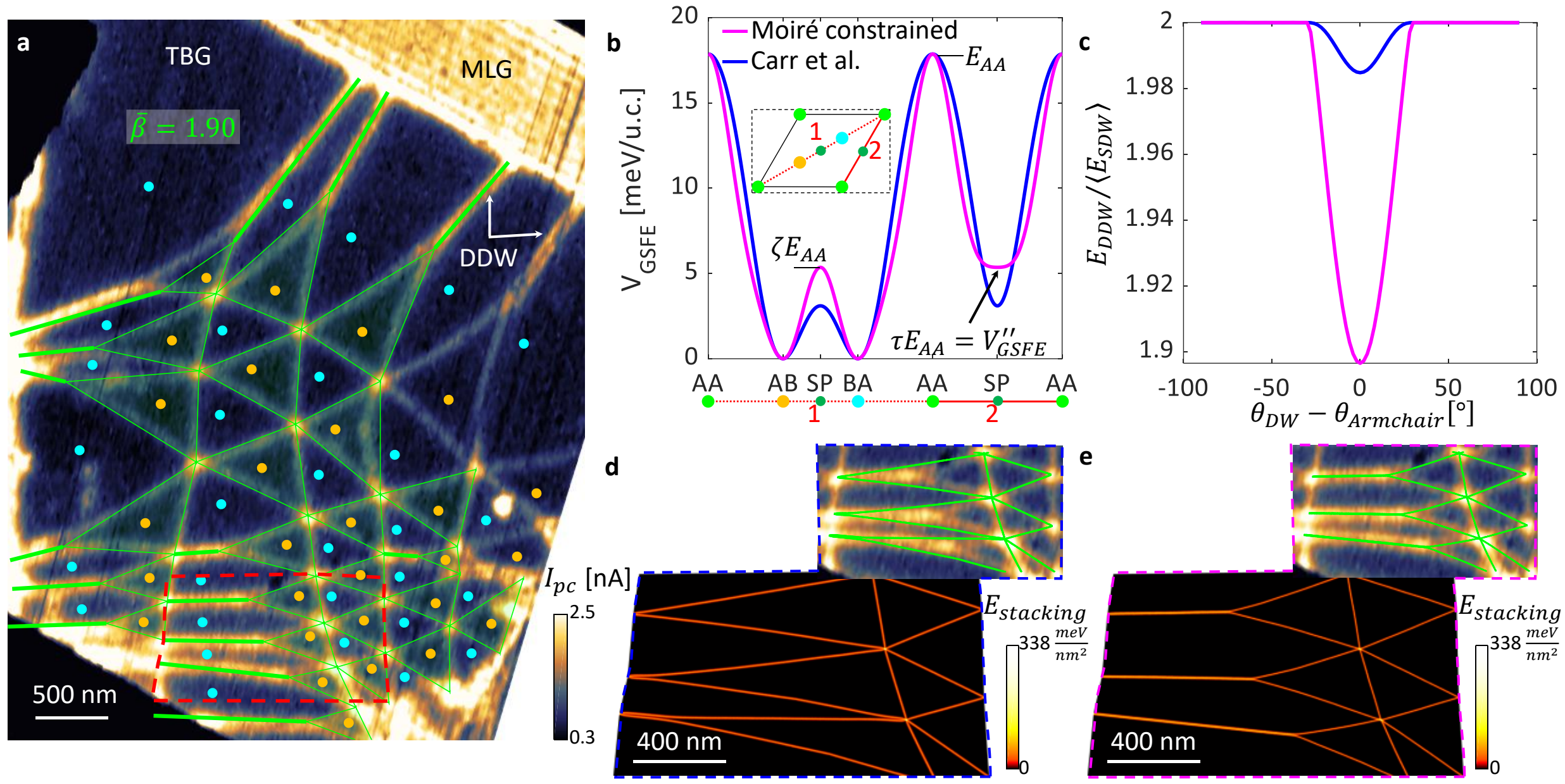

**Figure 2|Energy landscape of twisted bilayer graphene (TBG) revealed by the interplay between double (DDW) and single (SDW) domain walls**. **a,** Non-local nano-photocurrent map of moiré super-lattice of a TBG sample at the minimal twist limit (see methods for more details). The technique reveals the formation of DDWs (marked by "DDW") at strained domains, separating domains of identical stacking configurations (each configuration is indicated by dots of a given color [AB – orange, BA - cyan]). The green network overlaid on the data corresponds to the prediction by a single tuning parameter model with $\bar{\beta} = 1.90$ (see text and SI S6). **b,** GSFE of TBG from Ref. 31 (blue) and a moiré constrained version (magenta). The unit-less parameters $\zeta, \tau$, spanning the phase space of GSFE candidates for TBG, are illustrated (see SI S6). Inset: path in configuration space for presented GSFE line-cuts. **c,** Effective attraction between SDWs as reflected by DDW to SDW energy ratios for different SDW orientations (relative to armchair direction) for both models. $E_{DDW}$ is the DDW line-energy for a DDW along the armchair direction and similarly $\langle E_{SDW} \rangle$ is for the average of the two SDWs comprising the DDW. **d-e,** Stacking energy density from 2D relaxation calculation (see methods and SI S2) for the two discussed GSFE choices (**d** – literature, **e** – moiré constrained version with $\tau = 0.025, \zeta = 0.3$) of region marked by red dashed frame in **a**, showing fundamental differences in formation of DDWs. Inset: extracted domain wall structures from relaxation calculations overlaid on experimental results.

## Moiré metrology of twisted double bilayer graphene

Compared to TBG, the TDBG system makes an even more interesting case-study due to the small yet finite imbalance between the two lowest energy phases: Bernal (ABAB) and rhombohedral (ABCA) stackings[17]. This imbalance results in an energy cost per-unit-area ($\sigma$) for rhombohedral relative to Bernal stackings, leading to characteristic curved domains[17,38] (see Fig. 1a-b). Exploring large areas of TDBG reveals a rich distribution of rhombohedral domain shapes (see Fig. 1c and other TDBG images in this work). Figure 3a summarizes this distribution as a histogram of inverse curvature values ($\kappa^{-1}$), extracted from images as in Fig 1b (see SI section S6 for more examples). The histogram reveals a distinct clustering about a value of $\kappa^{-1} = 440 \pm 120\, nm$, which we use to assess the accuracy of several variants of the GSFE from available DFT functionals (Fig. 3b and see methods). All reported GSFE variants are qualitatively similar to the TBG case, peaking at the BAAC configuration, and having a saddle point barrier between ABAB and ABCA. A closer inspection (inset) reveals a profound difference between the GSFEs for the ABCA relative to the ABAB that governs domain curvature. We model the domain curvature and structure by a continuous 2D relaxation code (SI section S1). Figs. 3c-d show two representative cases, with disparate outcomes. In Fig. 3c (resembling the experimental case of Fig. 1b) the energy is minimized by slight bending of the SDW into the ABCA region. As the twist angle decreases (or as strain increases as in Fig. 1c), at some point it becomes energetically beneficial to form DDWs (Fig. 3d). As the twist angle further decreases, the shape of the ABCA domains remains unchanged. Similarly, solving for the domain formation for all DFT approaches and across a wide twist angle range we compare the extracted $\kappa^{-1}$. Interestingly, $\kappa$ is independent of the twist angle for all GSFE variants (with values indicated by colored lines over Fig. 3a), which is not generally the case (see discussion in SI S7) . The domain structures are further captured by the 2D "soap-bubble" model, as seen in turquoise dashed lines in the representative cases of Fig. 3c-d and more generally in Supplementary Videos 2-5. This model approximates the total energy as a sum of a domain area term and two line-energy terms as $E = \int_{SDW} dl\gamma_1(\varphi) + \int_{DDW} dl\gamma_2(\varphi) + \sigma S$, where $\gamma_{1,2}$ are the line-energies of SDW and DDW as a function of the domain-wall orientation respectively, the integrations are along the domain walls, and $S$ is the area of the domain (see SI section S7). All model parameters require only the GSFE (and elastic properties) to describe domain shapes, with no additional tuning parameters (SI section S7). One approach, DFT-D2, remarkably reproduces the experimental cluster (Fig. 3a), due to relatively high $\sigma$ and comparable line-energies to other approaches (see SI S2). It is noteworthy that the DFT-D approach has not been previously considered as the leading approach when theoretically benchmarked against the Quantum Monte-Carlo method for the binding energy of AA and AB stacking of bilayer graphene[39].

The rhombohedral domains represented in the histogram of Fig. 3a exemplify well-defined electrostatic environment near charge neutrality point (CNP) in the absence of the interlayer bias. As shown recently[38], upon charging and biasing the balance between the rhombohedral and Bernal phases can shift. An extreme demonstration of malleability of TDBG moiré patterns under a non-uniform distribution of charges and high strain conditions is presented in Fig. 3e. 3 holes (marked by blue circles) punctured one of the bilayers. This procedure prompts a highly strained moiré pattern, most strongly manifested in the densely packed parallel DDWs structures connecting the two bottom holes. The stack shows strong defect-induced doping (see discussion in SI S5), apparent in the enhanced nearfield contrast between the ABCA (dark) and ABAB (bright) phases (compare to contrast of Fig. 1c). Further support for the high non-uniformity of charge distribution is an observed region of flipped balance, where the ABAB phase becomes unstable relative to ABCA across a sharp (~50 nm) interface (to the left of the top hole). Attempting to model the moiré superlattice with the DFT-D2 GSFE at CNP (red in Fig. 3b) fails to capture the observed structure of excessively curved SDWs (color-map of Fig. 3f). However, when introducing a doping level of $8 \cdot 10^{12}\ cm^{-2}$ the resulting GSFE (dashed light-green in Fig. 3b) better captures the observed structure (green dots in Fig. 3f tracking the domain walls in the calculation). The difference between the two models becomes more pronounced for regions of higher strain, as highlighted in the inset of Fig. 3f (compare green and red dots in respect to the experimental map). Therefore, minuscule energy differences between models of order $0.1\ meV/u.c.$ (inset of Fig. 3b) result in measurable spatial features of the relaxed moiré patterns. To put this figure in context, the theoretical method which is widely considered as the gold-standard of ab-initio quantum chemistry[36,37] yields an accuracy as low as $3\ meV/atom$[37].

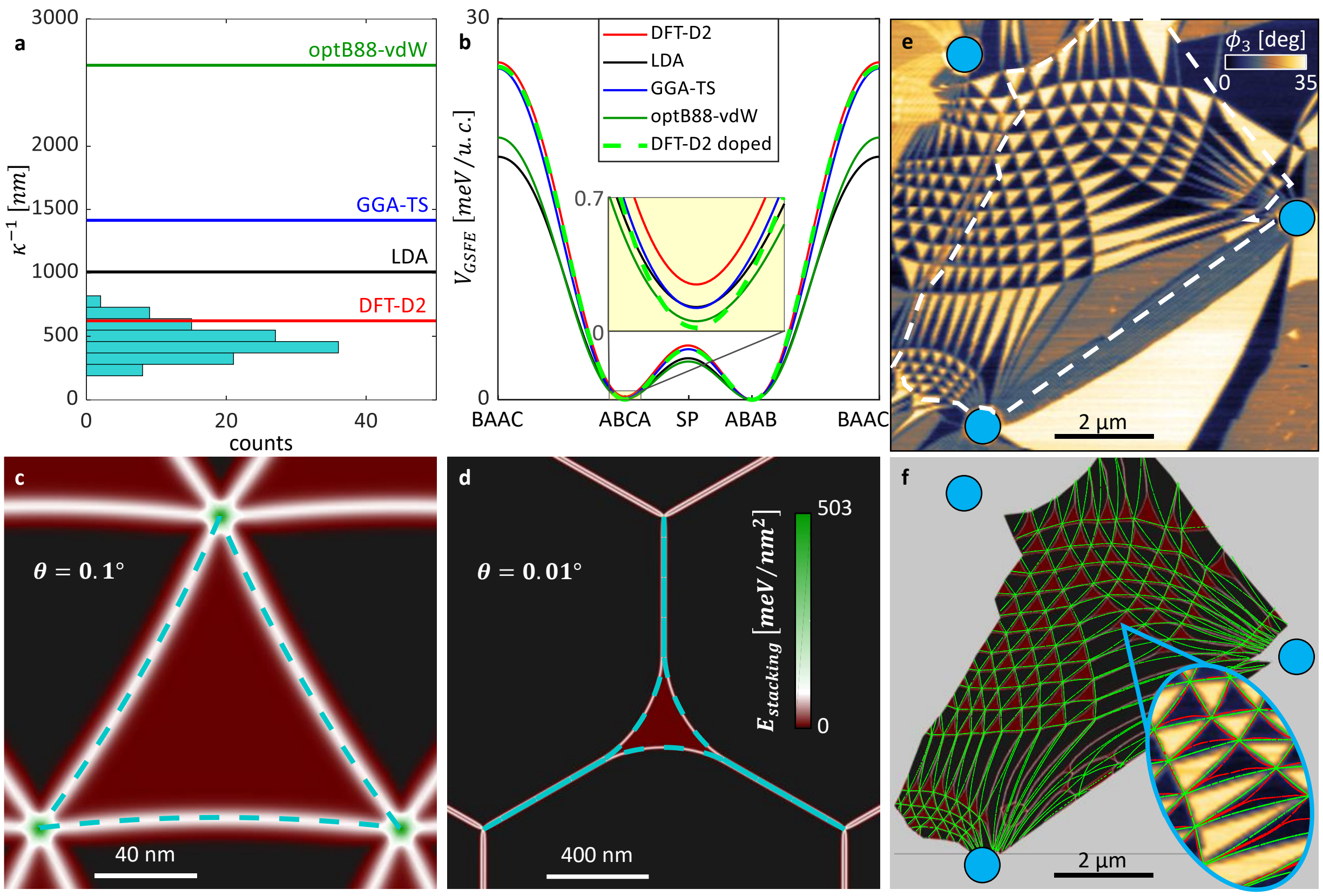

**Figure 3|Moiré super-lattice study of rhombohedral domains in twisted double bilayer graphene (TDBG). a,** Curvature histogram across all measured domains near charge neutrality point (see discussion in SI S7), showing a clear cluster at $440 \pm 120$ nm. Calculated curvatures of 4 DFT approaches (of **b**) are illustrated over the histogram (colored horizontal lines). **b,** GSFE of TDBG based on 4 different approaches (solid lines, see methods). The dashed light-green line is the GSFE for DFT-D2 approach at a doping level of $8 \cdot 10^{12}\ cm^{-2}$. Inset: enlarged view highlighting small difference between ABCA and ABAB configurations ($V_{GSFE}(ABAB) = 0$ identically). **c-d,** Mechanical relaxation solutions (false-color: stacking energy density) and "soap-bubble" model domain shape (dashed turquoise) for 2 representative twist angles (**c**: $\theta = 0.1°$, **d**: $\theta = 0.01°$) for DFT-D2. **e,** Mid-IR (940 $\mathrm{cm}^{-1}$) nearfield phase imaging of a defect-induced doped TDBG (see SI S7). 3 holes punctured one of the bilayers (blue circles - see methods) and induce a non-trivial external strain map. Mid-IR imaging resolves ABCA (dark) and ABAB (bright) phases and double-domain wall (DDW) formations (for instance, the multiple-DDW formation connecting bottom holes). **f,** Comparing relaxation calculations solutions of un-doped vs. $8 \cdot 10^{12}\ cm^{-2}$ doped DFT-D2 approach GSFE, simulating the experimental case (marked by dashed shape in **e),** False-color represents stacking energy density of un-doped case, overlaid (green dots) with tracked domain walls in the doped case. Inset: Highlighting differences between model by overlaying domain wall formation of doped (red) and un-doped (green) cases over strained region in the experimental map. The color-map is shared for **c-e.**

# Discussion

To understand the enhanced sensitivity of moiré metrology under strain (as seen in fig. 3f), we propose an alternative description of the moiré superlattice in terms of geometric interference pattern of the lattices of the two layers (see SI section S1). At minute twist angles, the relaxed moiré patterns are essentially a projection of the detailed energy landscape over space, accumulated over large regions compared to the atomic scale. The introduction of strain between the layers, whether naturally occurring or externally controlled, alters the interference pattern (SI section S8). As strain pushes the domain walls in Figs. 1-3 together, it also promotes their interaction; both effects are reflected in the relaxed moiré pattern (also see Supplementary Video 6).

Moiré metrology, introduced here, correlates first principle calculations of the stacking energy function with measurable spatial features of twisted vdW systems. The stacking energy function is widely used for modelling twisted multilayers across a broad range of twist angles and strain conditions, and has direct implications for the electronic band-structure[40]. Moiré metrology is not restricted to the discussed material systems and can be universally applied with other systems of great current interest[41–48]. Therefore, by providing a reliable account of the stacking energy function, moiré metrology has a broad impact across the field of vdW heterostructures. Furthermore, the moiré metrology tools can also be used for modelling and designing non-uniform strain fields in realistic devices. Finally, due to its outstanding stacking energy sensitivity, we propose moiré metrology as a concrete experimental path to provide much needed benchmarks for first-principle theoretical approaches[49].

# Methods

## Samples preparation

**Source crystals:** The bulk crystals of $MoSe_2$ and $WSe_2$ were grown by the self-flux method[50]: Single crystals of $MoSe_2$ were prepared by combining Mo powder (99.997%, Alfa Aesar 39686) and Se shot (99.999%, Alfa Aesar 10603) in a ratio of 1:50 (Mo:Se) in a quartz ampoule. The ampoules were subsequently sealed under vacuum ($\sim 5 \cdot 10^{-6}$ Torr). The reagents were then heated to 1000 ℃ within 24 hours and dwelled at this temperature for 8 weeks before being cooled to 350 ℃ over 4 weeks. At 350 ℃ the Se flux was decanted through alumina wool (Zircar D9202) in a centrifuge and the ampoules were quenched in air. The subsequently obtained $MoSe_2$ single crystals were annealed at 275 ℃ with a 200 ℃ gradient for 48h to remove any residual Se. A similar process was also used to synthesize single crystals of $WSe_2$ with 1:15 (W:Se) as the starting ratio and using W powder (99.999%, Alfa Aesar 12973). Kish graphite source crystals were purchased from Graphene Supermarket.

**Exfoliation:** The $MoSe_2$ and $WSe_2$ monolayers, and Graphene and hBN flakes were mechanically exfoliated from the bulk single crystals onto $SiO_2$/Si (285 nm oxide thickness) chips using the tape-assisted exfoliation technique (the tape used was Scotch Magic Tape). The exfoliation followed Ref. 51, such that the Si chips were treated with $O_2$ plasma (using a benchtop radio frequency oxygen plasma cleaner of Plasma Etch Inc., PE-50 XL, 100 W at a chamber pressure of ~215 mTorr) for 20 sec for graphene, for 10 sec for $MoSe_2$ and $WSe_2$, and no $O_2$ plasma treatment for hBN. The chips were then matched with respective exfoliation tape. In the graphene case the chip+tape assembly were heated at 100C for 60 sec and cooled to room temperature prior to removing the tape. Such thermal treatment was not done for other materials. The $MoSe_2$ and $WSe_2$ monolayer relative crystallographic orientation was obtained by linear-polarization- resolved second-harmonic generation (SHG).

**Stack preparation:** All heterostructures were assembled using standard dry-transfer techniques[52] with a polypropylene carbonate (PPC) film mounted on a transparent-tape-covered polydimethylsiloxane (PDMS) stamp. The transparent tape layer was added to the stamp to mold the PDMS into a hemispherical shape which provides precise control of the PPC contact area during assembly[53].

All graphene heterostructures were made by first picking up the bottom-layer boron nitride (h-BN) (> 25 nm thick), followed in the case of the TDBG samples by a graphite bottom gate-layer (> 5 nm thick), then a dielectric BN layer (> 25 nm thick). Prior to pick-up, mechanically exfoliated graphene flakes on $Si/SiO_2$ were separately patterned with anodic-oxidation lithography[54] to facilitate the "cut-and-stack" technique[55] where it was used (all samples but those used for STM, which used the established tearing method). In the case of the TDBG samples, additional anodic-oxidation lithography was used prior to pick-up to provide additional texture to the strain landscape, e.g. cut holes inside the bulk of one of the graphene layers or non-rectilinear edge geometries.

In the case of $MoSe_2/WSe_2$ the PPC was used to pick up a thin layer of exfoliated h-BN and a few layers of graphene. Then a monolayer $WSe_2$ was picked up and using the SHG data $MoSe_2$ monolayer was lifted on a rotation stage with ~1° twist angle. The stack was flipped over $Si/SiO_2$ (285nm) chip at 120°C. In the last step, the sample was thermally annealed in a high vacuum chamber to remove the PPC at 250°C for 1 hour.

Some of the presented measurements (TDBG stacks of Fig. 3e, Figs. S8b,d-f and red bins in histogram of S8a) were taken at this point while the stack was on a PDMS/transparent-tape/PPC structure. This provided access to the meta-stable large rhombohedral domain before they were suppressed by thermal annealing (see SI S9).

After optional mid-assembly scanning probe measurement and/or optional encapsulation of the twisted-graphene layers, the PPC film with the heterostructure on top is mechanically removed from the transparent-tape-covered PDMS stamp and placed onto a $Si/SiO_2$ substrate such that the final pick-up layer is the top layer.

In the case of the TBG device presented in this work, the underlying PPC was removed by vacuum annealing at T = 350˚ C. Standard plasma etching and metal deposition techniques[52] were then used to shape and make contact to the samples. In the case of the TDBG devices for STM (Fig. 1b) the stack was made using the established tearing method, using PPC as a polymer to sequentially pick up hBN, half of a piece of graphene followed by the second half with a twist angle.

For all samples for STM imaging, standard metal deposition techniques were avoided in order to maintain a pristine surface, therefore direct contact was made to the stack by micro-soldering with Field's metal[56], keeping temperatures below 80 C during the entire process.

## Nearfield imaging techniques

In this work we used two nano-optical imaging techniques: cryogenic nano-photocurrent imaging (used for TBG imaging) and phase-resolved scattering type scanning optical microscope imaging (used for TDBG imaging) (s-SNOM). Cryogenic photocurrent imaging[57] was done with a home-built cryogenic SNOM, and s-SNOM nearfield imaging was done with a commercial (Neaspec) SNOM[11]. In both cases using mid-IR light (continuous wave CO2 gas laser [Access Laser] at a wavelength of 10.6 µm) focused to a diffraction limited spot at the apex of a metallic tip, while raster scanning the sample at tapping mode. Fig. 2a was acquired at a temperature of 100 K and while tuning the silicon back-gate to a relatively high doping of $3 \cdot 10^{12}\ cm^{-2}$. In such a case we observe, for the first time, a non-local photo-current generation regime. In this regime, the light induced temperature profile is broad (relative to system size) and the photo-current generation is located at a distant interface (a monolayer twisted bilayer interface in this case, clearly visible as the bright region with plasmonic fringes at the top right section of Fig. 2a). The signal contrast in such a case results from absorption contrast between different stacking configuration, and not from thermo-electric properties. This unique approach provides a high-resolution image, not limited by thermal length-scales (see SI section S4 for more details).

In the s-SNOM case we collect the scattered light (power of 3 mW) by a cryogenic HgCdTe detector (Kolmar Technologies). The far-field contribution to the signal can be eliminated from the signal by locking to a high harmonic (here we used the 3rd harmonic of the tapping). The phase of the backscattered signal was extracted using an interferometric detection method, the pseudo-heterodyne scheme, by interfering the scattered light with a modulated reference arm at the detector. Fig. 1c was acquired with a level of $1 \cdot 10^{12}\ cm^{-2}$ p-doping applied with a Si backgate. Such a level of doping has negligible effect on domain curvature (see SI S5 for more details).

## STM and STS imaging

STM and STS measurements were carried out in a home-built STM under ultra-high vacuum conditions. $MoSe_2/WSe_2$ measurements were performed at 300 K while TDBG measurements were performed at 5.7 K. The setpoints of the STM (constant current mode) imaging were V= -1.8V and I=100pA for $MoSe_2/WSe_2$ measurements and V=0.5 V and I=50 pA for TDBG measurements. The tip-sample bias for the STS measurement of Fig. 1b was -57.5 meV. STM tips were prepared and calibrated for atomic sharpness and electronic integrity on freshly prepared Au (111) crystals. Samples were measured with multiple tips to ensure consistency of results.

## Parameter DFT calculations

The DFT calculations of the GSFE parameters were performed with the Vienna Ab initio Simulation Package (VASP)[58]. Plane wave basis sets were employed with energy cutoff of 1200 eV and 500 eV for the calculations for TDBG and $MoSe_2/WSe_2$, respectively. Pseudopotentials were constructed with the projector augmented wave method (PAW)[59,60]. $60 \times 60 \times 1$ and $11 \times 11 \times 1$ $\Gamma$-centered k-point grids were used in the calculations for TDBG and $MoSe_2/WSe_2$, respectively. Vacuum spacing larger than 15 Å was added along the z direction to eliminate the artificial interaction between periodic slab images in all the calculations. In the calculations of the GSFE, the x-y coordinates of the atoms in all the 2D layers were fixed and the z coordinates were allowed to relax until forces in the z-direction are less than 1 meV/Å for TDBG and 20 meV/Å for $MoSe_2/WSe_2$. The van der Waals interactions are important in evaluating the energetics of the 2D layer structures. We tested this effect for TDBG by considering four approaches: (1) Employing PAW pseudopotentials with the exchange-correlation functionals treated at the local density approximation (LDA) level. (2) Employing PAW pseudopotentials with the exchange-correlation functionals treated at the generalized gradient approximation (GGA) level[61] and additional van der Waals corrections are applied with the DFT-D2 method of Grimme[62]. (3) Employing PAW pseudopotentials with the GGA functionals and additional van der Waals corrections with the Tkatchenko-Scheffler method[63]. (4) Employing PAW pseudopotentials with a non-local correlation functional (optB88-vdW)[64–66] that approximately accounts for dispersion interactions. In the calculations for $MoSe_2/WSe_2$, we employed PAW pseudopotentials with the GGA PBE functionals with the DFT-D3 van der Waals corrections[67]. The bulk modulus and shear modulus for each material were calculated by applying isotropic or uniaxial strain to a monolayer lattice and then performing a quadratic fit to the strain-dependent energies. For TDBG, we assume the elastic coefficients are twice the values extracted for monolayer graphene. The DFT ground state energy is evaluated on a regular grid of different interlayer configurations. The Fourier components of the resulting energies are then extracted to create a convenient functional form for the GSFE.

## Data availability

The raw datasets used for the presented analysis within the current study are available from the corresponding author on reasonable request.

## Code availability

Developed relaxation codes can be provided from the corresponding author on reasonable request.

## References

1. Lopes Dos Santos, J. M. B., Peres, N. M. R. & Castro Neto, A. H. Graphene bilayer with a twist: Electronic structure. *Phys. Rev. Lett.* **99**, 19–22 (2007).

2. Bistritzer, R. & MacDonald, A. H. Moiré bands in twisted double-layer graphene. *Proc. Natl. Acad. Sci. U. S. A.* **108**, 12233–12237 (2011).

3. Yankowitz, M. *et al.* Tuning superconductivity in twisted bilayer graphene. *Science* **363**, 1059–1064 (2019).

4. Lu, X. *et al.* Superconductors, orbital magnets and correlated states in magic-angle bilayer graphene. *Nature* **574**, 653–657 (2019).

5. Sharpe, A. L. *et al.* Emergent ferromagnetism near three-quarters filling in twisted bilayer graphene. *Science* **365**, 605–608 (2019).

6. Zondiner, U. *et al.* Cascade of phase transitions and Dirac revivals in magic-angle graphene. *Nature* **582**, 203–208 (2020).

7. Stepanov, P. *et al.* Untying the insulating and superconducting orders in magic-angle graphene. *Nature* **583**, 375–378 (2020).

8. Uri, A. *et al.* Mapping the twist-angle disorder and Landau levels in magic-angle graphene. *Nature* **581**, 47–52 (2020).

9. Alden, J. S. *et al.* Strain solitons and topological defects in bilayer graphene. *Proc. Natl. Acad. Sci.* **110**, 11256–11260 (2013).

10. Cao, Y. *et al.* Unconventional superconductivity in magic-angle graphene superlattices. *Nature* **556**, 43–50 (2018).

11. Sunku, S. S. *et al.* Photonic crystals for nano-light in moiré graphene superlattices. *Science* **362**, 1153–1156 (2018).

12. Huang, S. *et al.* Topologically Protected Helical States in Minimally Twisted Bilayer Graphene. *Phys. Rev. Lett.* **121**, 37702 (2018).

13. Rickhaus, P. *et al.* Transport Through a Network of Topological Channels in Twisted Bilayer

Graphene. *Nano Lett.* **18**, 6725–6730 (2018).

14. Xu, S. G. *et al.* Giant oscillations in a triangular network of one-dimensional states in marginally twisted graphene. *Nat. Commun.* **10**, 1–5 (2019).

15. Hesp, N. C. H. *et al.* Collective excitations in twisted bilayer graphene close to the magic angle. *arXiv Prepr.* **1910.07893**, (2019).

16. Choi, Y. *et al.* Electronic correlations in twisted bilayer graphene near the magic angle. *Nat. Phys.* **15**, 1174–1180 (2019).

17. Kerelsky, A. *et al.* Moire-less Correlations in ABCA Graphene. *arXiv Prepr.* **1911.00007**, (2019).

18. Shen, C. *et al.* Correlated states in twisted double bilayer graphene. *Nat. Phys.* **16**, 520–525 (2020).

19. Liu, X. *et al.* Tunable spin-polarized correlated states in twisted double bilayer graphene. *Nature* **583**, 221–225 (2020).

20. Scuri, G. *et al.* Electrically Tunable Valley Dynamics in Twisted WSe2/WSe2 Bilayers. *Phys. Rev. Lett.* **124**, 217403 (2020).

21. Weston, A. *et al.* Atomic reconstruction in twisted bilayers of transition metal dichalcogenides. *Nat. Nanotechnol.* **15**, 592–597 (2020).

22. Rosenberger, M. R. *et al.* Twist Angle-Dependent Atomic Reconstruction and Moiré Patterns in Transition Metal Dichalcogenide Heterostructures. *ACS Nano* **14**, 4550–4558 (2020).

23. Wu, F., Lovorn, T., Tutuc, E. & Macdonald, A. H. Hubbard Model Physics in Transition Metal Dichalcogenide Moiré Bands. *Phys. Rev. Lett.* **121**, 26402 (2018).

24. Yu, H., Liu, G. Bin, Tang, J., Xu, X. & Yao, W. Moiré excitons: From programmable quantum emitter arrays to spin-orbit–coupled artificial lattices. *Sci. Adv.* **3**, e1701696 (2017).

25. Wang, J. *et al.* Diffusivity Reveals Three Distinct Phases of Interlayer Excitons in MoSe2/WSe2 Heterobilayers. *arXiv e-prints* 2001.03812 (2020).

26. Shimazaki, Y. *et al.* Strongly correlated electrons and hybrid excitons in a moiré heterostructure. *Nature* **580**, 472–477 (2020).

27. Jin, C. *et al.* Observation of moiré excitons in WSe2/WS2 heterostructure superlattices. *Nature* **567**, 76–80 (2019).

28. Balents, L., Dean, C. R., Efetov, D. K. & Young, A. F. Superconductivity and strong correlations in moiré flat bands. *Nat. Phys.* **16**, 725–733 (2020).

29. Kennes, D. M. *et al.* Moiré heterostructures : a condensed matter quantum simulator. *Nat. Phys.* 1–12 (2020).

30. Yankowitz, M., Ma, Q., Jarillo-Herrero, P. & LeRoy, B. J. van der Waals heterostructures combining graphene and hexagonal boron nitride. *Nat. Rev. Phys.* **1**, 112–125 (2019).

31. Carr, S. *et al.* Relaxation and domain formation in incommensurate two-dimensional heterostructures. *Phys. Rev. B* **98**, 224102 (2018).

32. Cazeaux, P., Luskin, M. & Massatt, D. Energy Minimization of Two Dimensional Incommensurate

Heterostructures. *Arch. Ration. Mech. Anal.* **235**, 1289–1325 (2020).

33. Enaldiev, V. V., Zólyomi, V., Yelgel, C., Magorrian, S. J. & Fal'ko, V. I. Stacking Domains and Dislocation Networks in Marginally Twisted Bilayers of Transition Metal Dichalcogenides. *Phys. Rev. Lett.* **124**, 206101 (2020).

34. Zhou, S., Han, J., Dai, S., Sun, J. & Srolovitz, D. J. Van der Waals bilayer energetics: Generalized stacking-fault energy of graphene, boron nitride, and graphene/boron nitride bilayers. *Phys. Rev. B* **92**, 155438 (2015).

35. Wang, W. *et al.* Measurement of the cleavage energy of graphite. *Nat. Commun.* **6**, 7853 (2015).

36. Dykstra, C. E., Frenking, G., Kim, K. S. & Scuseria, G. E. Computing technologies, theories, and algorithms: The making of 40 years and more of theoretical and computational chemistry. *Theory Appl. Comput. Chem.* 1–7 (2005). doi:10.1016/B978-044451719-7/50044-5

37. Neese, F., Atanasov, M., Bistoni, G., Maganas, D. & Ye, S. Chemistry and Quantum Mechanics in 2019: Give Us Insight and Numbers. *J. Am. Chem. Soc.* **141**, 2814–2824 (2019).

38. Li, H. *et al.* Global Control of Stacking-Order Phase Transition by Doping and Electric Field in Few-Layer Graphene. *Nano Lett.* **20**, 3106–3112 (2020).

39. Mostaani, E., Drummond, N. D. & Fal'Ko, V. I. Quantum Monte Carlo Calculation of the Binding Energy of Bilayer Graphene. *Phys. Rev. Lett.* **115**, 115501 (2015).

40. Carr, S., Fang, S., Zhu, Z. & Kaxiras, E. Exact continuum model for low-energy electronic states of twisted bilayer graphene. *Phys. Rev. Res.* **1**, 013001 (2019).

41. Tong, Q., Liu, F., Xiao, J. & Yao, W. Skyrmions in the Moiré of van der Waals 2D Magnets. *Nano Lett.* **18**, 7194–7199 (2018).

42. Van Wijk, M. M., Schuring, A., Katsnelson, M. I. & Fasolino, A. Moiré patterns as a probe of interplanar interactions for graphene on h-BN. *Phys. Rev. Lett.* **113**, 135504 (2014).

43. Woods, C. R. *et al.* Charge-polarized interfacial superlattices in marginally twisted hexagonal boron nitride. *arXiv e-prints* 2010.06914 (2020).

44. Yasuda, K., Wang, X., Watanabe, K., Taniguchi, T. & Jarillo-Herrero, P. Stacking-engineered ferroelectricity in bilayer boron nitride. *arXiv e-prints* 2010.06600 (2020).

45. Woods, C. R. *et al.* Commensurate-incommensurate transition in graphene on hexagonal boron nitride. *Nat. Phys.* **10**, 451–456 (2014).

46. Chen, X. *et al.* Moiré engineering of electronic phenomena in correlated oxides. *Nat. Phys.* **16**, 631–635 (2020).

47. Xian, L., Kennes, D. M., Tancogne-Dejean, N., Altarelli, M. & Rubio, A. Multiflat Bands and Strong Correlations in Twisted Bilayer Boron Nitride: Doping-Induced Correlated Insulator and Superconductor. *Nano Lett.* **19**, 4934–4940 (2019).

48. Kennes, D. M., Xian, L., Claassen, M. & Rubio, A. One-dimensional flat bands in twisted bilayer germanium selenide. *Nat. Commun.* **11**, 1124 (2020).

49. Mata, R. A. & Suhm, M. A. Benchmarking Quantum Chemical Methods: Are We Heading in the Right Direction? *Angew. Chemie - Int. Ed.* **56**, 11011–11018 (2017).

50. Edelberg, D. *et al.* Approaching the Intrinsic Limit in Transition Metal Diselenides via Point Defect Control. *Nano Lett.* **19**, 4371–4379 (2019).

51. Huang, Y. *et al.* Reliable Exfoliation of Large-Area High-Quality Flakes of Graphene and Other Two-Dimensional Materials. *ACS Nano* **9**, 10612–10620 (2015).

52. Wang, L. *et al.* One-dimensional electrical contact to a two-dimensional material. *Science* **342**, 614–617 (2013).

53. Kim, K. *et al.* Van der Waals Heterostructures with High Accuracy Rotational Alignment. *Nano Lett.* **16**, 1989–1995 (2016).

54. Li, H. *et al.* Electrode-Free Anodic Oxidation Nanolithography of Low-Dimensional Materials. *Nano Lett.* **18**, 8011–8015 (2018).

55. Saito, Y., Ge, J., Watanabe, K., Taniguchi, T. & Young, A. F. Independent superconductors and correlated insulators in twisted bilayer graphene. *Nat. Phys.* **16**, 926–930 (2020).

56. Girit, C. O. & Zettl, A. Soldering to a single atomic layer. *Appl. Phys. Lett.* **91**, 193512 (2007).

57. Sunku, S. S. *et al.* Nano-photocurrent Mapping of Local Electronic Structure in Twisted Bilayer Graphene. *Nano Lett.* **20**, 2958–2964 (2020).

58. Kresse, G. & Furthmüller, J. Efficient iterative schemes for ab initio total-energy calculations using a plane-wave basis set. *Phys. Rev. B* **54**, 11169–11186 (1996).

59. Blöchl, P. E. Projector augmented-wave method. *Phys. Rev. B* **50**, 17953–17979 (1994).

60. Kresse, G. & Joubert, D. From ultrasoft pseudopotentials to the projector augmented-wave method. *Phys. Rev. B* **59**, 1758–1775 (1999).

61. Perdew, J. P., Burke, K. & Ernzerhof, M. Generalized gradient approximation made simple. *Phys. Rev. Lett.* **77**, 3865–3868 (1996).

62. Grimme, S. Semiempirical GGA-Type Density Functional Constructed with a Long-Range Dispersion Correction. *J. Comput. Chem.* **27**, 1787–1799 (2006).

63. Tkatchenko, A. & Scheffler, M. Accurate molecular van der Waals interactions from ground-state electron density and free-atom reference data. *Phys. Rev. Lett.* **102**, 073005 (2009).

64. Dion, M., Rydberg, H., Schröder, E., Langreth, D. C. & Lundqvist, B. I. Van der Waals density functional for general geometries. *Phys. Rev. Lett.* **92**, 246401 (2004).

65. Klimeš, J., Bowler, D. R. & Michaelides, A. Van der Waals density functionals applied to solids. *Phys. Rev. B* **83**, 195131 (2011).

66. Klimeš, J., Bowler, D. R. & Michaelides, A. Chemical accuracy for the van der Waals density functional. *J. Phys. Condens. Matter* **22**, 022201 (2010).

67. Grimme, S., Antony, J., Ehrlich, S. & Krieg, H. A consistent and accurate ab initio parametrization of density functional dispersion correction (DFT-D) for the 94 elements H-Pu. *J. Chem. Phys.* **132**, 154104 (2010).

## Acknowledgements

Research at Columbia on moiré superlattice is supported as part of Programmable Quantum Materials, an Energy Frontier Research Center funded by the U.S. Department of Energy (DOE), Office of Science, Basic Energy Sciences (BES), under award DE-SC0019443. STM instrumentation for STM experiments was developed with support from the Air Force Office of Scientific Research via grant FA9550-16-1-0601. Synthesis of $MoSe_2$ and $WSe_2$ crystals was supported by the NSF MRSEC program through Columbia in the Center for Precision Assembly of Superstratic and Superatomic Solids (DMR-2011738). A.R. acknowledges support by the European Research Council (ERC-2015-AdG-694097), Grupos Consolidados (IT1249-19), SFB925 and the Flatiron Institute, a division of the Simons Foundation. We acknowledge funding by the Deutsche Forschungsgemeinschaft (DFG) under Germany's Excellence Strategy - Cluster of Excellence Matter and Light for Quantum Computing (ML4Q) EXC 2004/1 – 390534769, funding by Advanced Imaging of Matter (AIM) EXC 2056 – 390715994, funding by the DFG under RTG 1995 and RTG 2247 and by DFG within the Priority Program SPP 2244 "2DMP". We acknowledge support from the Max Planck-New York City Center for Non-Equilibrium Quantum Phenomena. Work at Harvard was supported by the STC Center for Integrated Quantum Materials NSF Grant No. DMR1231319 and ARO MURI Award No. W911NF-14-0247. D.H. was supported by a grant from the Simons Foundation (579913). N.R.F. acknowledges support from the Stewardship Science Graduate Fellowship program provided under cooperative agreement number DE-NA0003864. C.R.V. acknowledges funding from the European Union's Horizon 2020 research and innovation programme under the Marie Skłodowska-Curie grant agreement No 844271. D.N.B. is the Vannevar Bush Faculty Fellow ONR-VB: N00014-19-1-263 and the Moore Investigator in Quantum Materials EPIQS #9455.

## Author contributions

D.H. conceived the moiré metrology framework, developed the relaxation codes and performed related analysis. S.Chen and D.G.A. prepared the TBG device with supervision by J.C.H. and C.R.D. N.R.F. and C.Z. fabricated all TDBG devices for nano-optics experiments with supervision by J.C.H. and C.R.D. L.W. fabricated TDBG devices for STM experiments with supervision by J.C.H. S.S. fabricated T-H-$MoSe_2$/$WSe_2$ devices and performed their STM study with supervision by A.N.P. D.R. grew the $MoSe_2$ and $WSe_2$ crystals with supervision by J.C.H. K.W. and T.T. grew the hBN crystals. Cryogenic nano-photocurrent imaging of TBG was done by D.H. and S.S.S. with assistance from A.S.M. and supervision by D.N.B. D.H. performed s-SNOM imaging of TDBG with supervision by D.N.B. A.K. and C.R.V. performed and analyzed the STM/STS measurements of TDBG with supervision by A.N.P. L.X. performed DFT calculations for GSFE of TDBG with supervision by A.R. and S.Carr performed DFT calculations for GSFE of $MoSe_2$/$WSe_2$ with supervision by E.K. Manuscript was written by D.H. and D.N.B. with contributions by D.M.K., N.R.F., S.Carr, A.S.M. and A.R. All authors contributed by discussions with particular contribution with interpretation of theory by D.M.K, S.Carr and A.R.

## Competing interests

The authors declare no competing interests.

# Supplementary Information - Moiré metrology of energy landscapes in van der Waals heterostructures

Dorri Halbertal[1*], Nathan R. Finney[2], Sai S. Sunku[1], Alexander Kerelsky[1], Carmen Rubio-Verdú[1], Sara Shabani[1], Lede Xian[3,4], Stephen Carr[5,6], Shaowen Chen[1,7], Charles Zhang[1,8], Lei Wang[1], Derick Gonzalez-Acevedo[1,7], Alexander S. McLeod[1], Daniel Rhodes[1,9], Kenji Watanabe[10], Takashi Taniguchi[10], Efthimios Kaxiras[11], Cory R. Dean[1], James C. Hone[2], Abhay N. Pasupathy[1], Dante M. Kennes[3,12], Angel Rubio[3,13], D. N. Basov[1]

## S 1. Relaxation models

In this work two real space simulation tools have been developed for the solution of the relaxation problem, a 1D and a full 2D code. In both cases the energy functional depends on position, the displacement field of each layer and their spatial gradients, and is composed of an elastic energy term and a stacking energy term. These models assume that the inter-layer displacement field changes on a length-scale larger than $\alpha$, the atomic unit-cell spacing. As a result, the local stacking energy density is assumed to be a direct consequence of the generalized stacking fault energy function (GSFE), by assuming that the local stacking energy density is the same as for an infinite bulk arrangement of the same stacking configuration. The total energy would be $E = \int d^2r\mathcal{E}(\boldsymbol{r}, \boldsymbol{u}, \boldsymbol{\nabla u})$ where the energy functional in its 2D spatial coordinate form is:

$$\mathcal{E}(\boldsymbol{r}, \boldsymbol{u}, \boldsymbol{\nabla u}) = \mathcal{E}_{elastic}(\boldsymbol{\nabla u}) + \mathcal{E}_{stacking}(\boldsymbol{r}, \boldsymbol{u})$$

Where $\boldsymbol{u}$ is the inter-layer displacement, and we assume that each layer is displaced such that: $\boldsymbol{u_t} = \mu\boldsymbol{u}, \boldsymbol{u_b} = -(1-\mu)\boldsymbol{u}$ for a dimensionless parameter $\mu$ allowing to tune the layer motion of bottom (b) and top (t) layers, where $0 \le \mu \le 1$. $\mu = \frac{1}{2}$, for instance, is the case where the two layers are displaced in an anti-symmetric fashion. In the simulations presented in this work we used $\mu = \frac{1}{2}$. The conclusions of the TBG section of the paper do not depend on this choice. For TDBG this choice is justified in SI section S2 below.

[1]Department of Physics, Columbia University, New York, NY, USA.
[2]Department of Mechanical Engineering, Columbia University, New York, NY, USA.
[3]Max Planck Institute for the Structure and Dynamics of Matter and Center Free-Electron Laser Science, Luruper Chaussee 149, 22761 Hamburg, Germany.
[4]Present address: Songshan Lake Materials Laboratory, Dongguan, Guangdong 523808, China.
[5]Department of Physics, Harvard University, Cambridge, Massachusetts 02138, USA.
[6]Present address: Brown University, Providence, RI 02912, USA.
[7]Present address: Department of Physics, Harvard University, Cambridge, MA 02138, USA.
[8]Present address: Department of Physics, University of California at Santa Barbara, Santa Barbara, CA 93106, USA.
[9]Present address: Department of Materials Science and Engineering, University of Winsconsin-Madison, WI 53706, USA.
[10]National Institute for Material Science, Tsukuba, Japan
[11]John A. Paulson School of Engineering and Applied Sciences, Harvard University, Cambridge, Massachusetts 02138, USA.
[12]Institut fur Theorie der Statistischen Physik, RWTH Aachen University, 52056 Aachen, Germany.
[13]Center for Computational Quantum Physics, Flatiron Institute, New York, NY 10010 USA.
*Correspondence and requests for materials should be addressed to D.H. (dh2917@columbia.edu).

The elastic term has the explicit form:

$$\mathcal{E}_{elastic}(\nabla \boldsymbol{u}) = \frac{1}{2}K\left(\partial_x u_x + \partial_y u_y\right)^2 + \frac{1}{2}G\left(\left(\partial_x u_x - \partial_y u_y\right)^2 + \left(\partial_x u_y + \partial_y u_x\right)^2\right)$$
$$K = (1-\mu)^2 K_b + \mu^2 K_t, G = (1-\mu)^2 G_b + \mu^2 G_t$$

Where $K_t$ ($K_b$) and $G_t$ ($G_b$) are the bulk and shear elastic moduli of the top (bottom) layer.

The stacking energy term has the form:

$$\mathcal{E}_{stacking}(\boldsymbol{r}, \boldsymbol{u}) = V_{GSFE}\big(v(\boldsymbol{r}, \boldsymbol{u}), w(\boldsymbol{r}, \boldsymbol{u})\big)$$

Where the GSFE in its general form follows the notation of Ref. 31:

$$\begin{aligned} V_{GSFE}(v,w) = {} & c_0 + c_1(\cos v + \cos w + \cos(v+w)) + c_2(\cos(v+2w) + \cos(v-w) + \cos(2v+w)) \\ & + c_3(\cos 2v + \cos 2w + \cos(2v+2w)) + c_4(\sin v + \sin w - \sin(v+w)) \\ & + c_5(\sin(2v+2w) - \sin 2w - \sin 2v) \end{aligned}$$

And $\Omega \equiv \begin{pmatrix} v(\boldsymbol{r}, \boldsymbol{u}) \\ w(\boldsymbol{r}, \boldsymbol{u}) \end{pmatrix}$ tracks the stacking configuration of the two layers. We assume the two layers have a hexagonal lattice with Bravais spacing of $\alpha_b = \alpha, \alpha_t = (1+\delta)\alpha$, where $\delta$ is the lattice mismatch. We assume a global twist $\theta$ and displacement $\Delta\boldsymbol{r}$ between the layers, such that in absence of additional local displacements the top layer's AA sites will be located at $\frac{v}{2\pi}\boldsymbol{b}_1 + \frac{w}{2\pi}\boldsymbol{b}_2$ for $v, w$ which are integer multiples of $2\pi$ where $\boldsymbol{b}_1 = \begin{pmatrix} \cos\theta_0 \\ \sin\theta_0 \end{pmatrix}, \boldsymbol{b}_2 = \begin{pmatrix} \cos(\theta_0 + \frac{\pi}{3}) \\ \sin(\theta_0 + \frac{\pi}{3}) \end{pmatrix}$ are the normalized Bravais vectors of the bottom layer and $\theta_0$ is the orientation of the zigzag direction. For such definitions one can show that the local configuration would be:

$$\Omega(\boldsymbol{r}, \boldsymbol{u}) = \frac{2\pi}{\alpha}[\boldsymbol{b}_1 \quad \boldsymbol{b}_2]^{-1}\left((R_{-\theta} - (1+\delta)I_2)\boldsymbol{r} - \big(\mu R_{-\theta} + (1-\mu)(1+\delta)\big)\boldsymbol{u} - R_{-\theta}\Delta\boldsymbol{r}\right)$$

Where $R_{-\theta} = \begin{pmatrix} \cos\theta & \sin\theta \\ -\sin\theta & \cos\theta \end{pmatrix}$ is a rotation matrix by the global twist angle and $I_2$ is the 2D identity matrix. For the case of a homo-bilayer system and assuming a fixed bottom layer without global translation, the expression can be simplified to be:

$$\Omega(\boldsymbol{r}, \boldsymbol{u}) = \frac{2\pi}{\alpha}[\boldsymbol{b}_1 \quad \boldsymbol{b}_2]^{-1}\big((R_{-\theta} - I_2)\boldsymbol{r} - R_{-\theta}\boldsymbol{u}\big)$$

The remaining component for a well-defined problem is the boundary conditions (BC). For the 1D case, used for solving for the structure of single (SDW) and double (DDW) domain walls in the discussed systems, we assume $u_x$ and $u_y$ depend only on $x$ (which further simplifies the expressions), and assume a predefined boundary conditions for the stacking configurations on both sides of the domain wall dislocation. In such an approach we can describe both SDW (as an AB/BA interface) and DDW (as an interface between AB and AB across a BA segment in configuration space). From such calculations we can extract the energy costs per unit length angular functions $\gamma_1(\varphi)$ (for TBG) and $\gamma_2(\varphi)$ (for TBG and TDBG).

These functions have a $\pi$ periodicity, and can be approximated as $\gamma_j(\varphi) = \left(E_{S,j}^n \cos^2\varphi + E_{T,j}^n \sin^2\varphi\right)^{\frac{1}{n}}, j = SDW, DDW$ where $\varphi$ is the angle relative to the shear direction of the dislocation. Note that in the limit $n \to 0$, the expression converges to $\gamma_j(\varphi) = E_{S,j}\left(\frac{E_{T,j}}{E_{S,j}}\right)^{\sin^2\varphi}$. The fitted parameters, following such 1D calculations, which were used in the analysis throughout this work are detailed in table 1. The SDW in the case of TDBG is not a stable 1D soliton structure, therefore one needs a full 2D solution in order to extract $\gamma_1(\varphi)$ for TDBG as done in SI section S7 and summarized in Table 1 (SI section S2).

The boundary conditions in the 2D cases provided throughout the paper are more involved. They include rigid pre-defined stacking at selected points (as in Fig. 1f-g, Fig. 2d-e and Fig. 3f of the main text) and periodic boundary conditions (as in Fig. 1a,e and Figs. 3c-d of the main text). In the rigid BC case we impose the stacking configuration at selected points, while the periodic boundary conditions impose the periodicity of the moiré super-lattice in the displacement field, possibly with an additional uniform external strain.

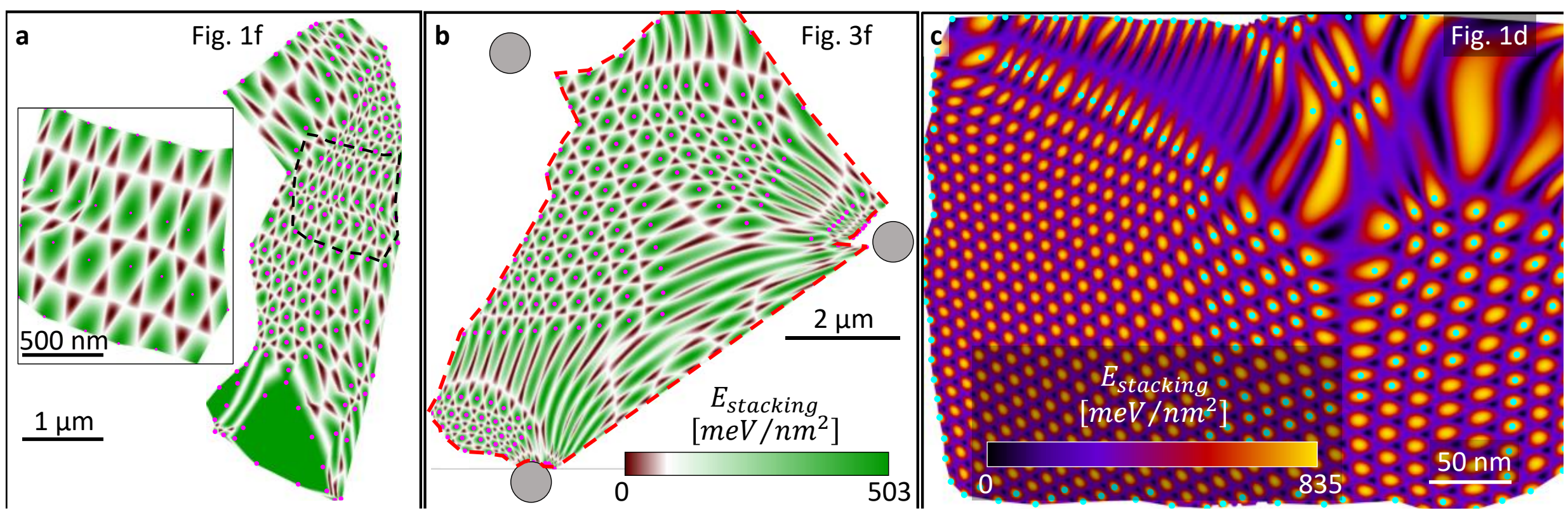

**Figure S1|Initial and boundary conditions for simulations in this work.** Colored dots mark points used to impose a given stacking configuration. False color map corresponds to stacking energy density resulting from the interpolation of the displacement field between these points. **a,** The TDBG case presented in Fig. 1c. Inset corresponds to modified initial conditions that warps the moiré dislocation from the other side (in region marked by dashed frame). Fig. 1f is composed of a combination of the two relaxation calculations. **b,** The TDBG case presented in Fig. 3f. **c,** The $MoSe_2/WSe_2$ case presented in Fig. 1g. **a** and **b** share a color-map.

The energy functional was minimized with standard unconstrained minimization techniques (trust region algorithm), using an interpolation between the BC points as an initial guess (as detailed in next section). The simulation tools used in this work were verified against calculations in Ref. 31, and with other results generated by the same code, which was courteously provided by the authors of Ref. 31.

## S 2. Details of relaxation calculations and material properties

For completeness, we provide details of 2D relaxation calculations performed throughout this work. Fig. 1a: The atomic positions in the sketch of Fig. 1a were calculated using a periodic relaxation calculation. The TDBG calculation (top) used the DFT-D2 model for a twist angle of 3° (moiré period of 4.72 nm), but with an interlayer coupling increased by a factor of $2 \cdot 10^4$, as to induce stronger curvature and formation of DDW. The T-H-$MoSe_2/WSe_2$ (bottom) used the GSFE detailed in Table 1 (Si section S2), with a twist angle of $3.989°$ (to have similar moiré period as in the TDBG case) and an interlayer coupling increased by a factor of $25$ (similarly motivated as in the TDBG case). Fig. 1e: For Fig. 1e a twist angle of 0.081° and an additional small external strain of 0.026% using a Poisson ratio of $\nu = 0.22$ was introduced, in order to model the experimental results of Fig. 1b. Fig. 1f, 1d, 3f, S9b: The simulation boundary and initial conditions (the initial guess used in the optimization process) for these calculations are shown in Figure S1. The used GSFE parameters for each material are listed in Table 1 (DFT-D2 approach for TDBG). For each simulation a given configuration was imposed at the points marked by circles. The false color maps represent the initial condition that was set by spline interpolation of the displacement field in between these points. Fig. 1f required a special treatment due to the existence of a dislocation in the moiré superlattice (see dashed circle in Fig. 1f). In order to account for the dislocation in the stacking energy density map, two calculations were performed one covered by the panel of Fig. S1a, and the other by the inset of Fig. S1a that covers the dislocation from the other side. Stitching the two images allow to have some account of the dislocation, even though the presented model does not implicitly support such cases. Fig. 2d-e are covered in SI section S4 below. Fig. 3c-d Fig. S6b and Supplementary Videos 2-5 assumed zero external strain, and were solved with periodic boundary conditions. Figs. S2d-f and Supplementary Video 6 were solved with periodic boundary conditions and a uniform external strain tensor as indicated in SI section S8. The calculation of Fig. S6a is discussed in detail in SI section S7.

| Material | TBG[31,34] | TBG - mod. $\tau = 0.025$ $\zeta = 0.3$ | TDBG (DFT-D2) | TDBG (LDA) | TDBG (GGA-TS) | TDBG (optB88-vdW) | $MoSe_2/WSe_2$ 180◦ twist |
|---|---|---|---|---|---|---|---|
| K | 69518 | 69518 | 139036 | 139036 | 139036 | 139036 | 40521 43113 |
| G | 47352 | 47352 | 94704 | 94704 | 94704 | 94704 | 26464 30770 |
| $c_0$ | 6.832 | 6.1891 | 10.4395 | 7.6484 | 9.7361 | 7.7155 | 42.6 |
| $c_1$ | 4.064 | 3.2041 | 6.0761 | 4.3773 | 5.861 | 4.6825 | 16.0 |
| $c_2$ | -0.374 | -0.07855 | -0.4995 | -0.4088 | -0.3330 | -0.2712 | -2.7 |
| $c_3$ | -0.095 | 0.76486 | -0.1972 | -0.1384 | -0.0771 | -0.0989 | -1.1 |
| $c_4$ | 0 | 0 | 0.0453 | 0.02196 | 0.0141 | 0.0041 | 3.7 |
| $c_5$ | 0 | 0 | 0.0019 | 0.0025 | 0.0093 | 0.0060 | 0.6 |
| $E_{S,SDW}$ | 953 | 1201 | 1103 | 965 | 1062 | 928 | N/A |
| $E_{T,SDW}$ | 1497 | 1887 | 1776 | 1531 | 1681 | 1460 | |
| $n_{SDW}$* | 1.49 | 1.04 | 2** | 2** | 2** | 2** | |
| $E_{S,DDW}$ | 2184 | 2606 | 2609 | 2267 | 2492 | 2171 | |
| $E_{T,DDW}$ | 2744 | 3433 | 3339 | 2890 | 3170 | 2747 | |
| $n_{DDW}$* | 0.72 | 4.16 | 1.43 | 1.34 | 1.266 | 1.01 | |
| $\sigma$ | 0 | 0 | 4.637 | 2.404 | 2.309 | 0.997 | |

**Table 1**: Material coefficients (above bold line) and extracted parameters (below bold line) used in this work. All material coefficients have units of $meV/u.c.$, all graphene heterostructures use a unit-cell (u.c.) spacing of $\alpha = 0.247\ nm$, while $\alpha = 0.3288\ nm$ for $MoSe_2$ and $\alpha = 0.3282\ nm$ for $WSe_2$. $K, G$ are the bulk and shear moduli (for $MoSe_2/WSe_2$ hetero-bilayer the values for each layer are mentioned separately), and $c_0 - c_5$ are the generalized stacking fault energy function (GSFE) coefficients, following the notation of Ref. 31. For TBG: $K, G, c_0 - c_5$ are taken from Ref. 31. "TBG – mod. $\tau = 0.025, \zeta = 0.3$" denotes the moiré constrained version of the parameters of TBG of Ref. 31 in order to reproduce experimental formations of DDWs (see Fig. 2 and discussion in SI section S4). The GSFE parameters of TDBG were calculated using different DFT approaches (methods for more details), while $K, G$ were simply taken as twice that of TBG. $E_{S,SDW}$, $E_{T,SDW}$, $E_{S,DDW}$ and $E_{T,DDW}$ are in units of $meV/nm$ and $\sigma$ in $meV/nm^2$. *Exponents are fitted to provide analytic expression approximating the calculated angular curves. **SDW parameters for TDBG were extracted from fitting an elliptic arc to the SDW as discussed in SI S7, therefore assuming $n_{SDW} = 2$.

In all calculation we used $\mu = 0.5$ (see SI section S1 above), that distributes the displacement equally between the two layers of the calculation. While this choice makes physical sense, and has been commonly used in literature as well[31], it is in fact supported by the experimental results presented in Fig. 3 for TDBG. Deviating from $\mu = 0.5$ makes the layers effectively stiffer, thus increasing the line energies of a domain wall. Since the ABCA stacking energy density dependence only on the GSFE, changing $\mu$ away from this value will lead to an increase in $\kappa^{-1}$. Therefore, the comparison of the histogram of Fig. 3a with relaxation calculations of the 4 explored DFT approaches supports the choice $\mu = 0.5$.

The material properties that were used in the relaxation calculations are summarized in Table 1 (of different DFT approaches from literature where referenced, and otherwise calculated within this work). Relevant extracted parameters are also mentioned (below the bold lines).

## S 3. Evaluation of the generalized stacking fault energy function of MoSe2/WSe2

The map of Fig. 1d reveals the formation of double domain walls (DDWs) in strained region of the twisted H-stacked $MoSe_2/WSe_2$ structure. Here we compare the experimental DDW structure with the relaxation solution using calculated generalized stacking fault energy function (GSFE – see Table 1). Fig.

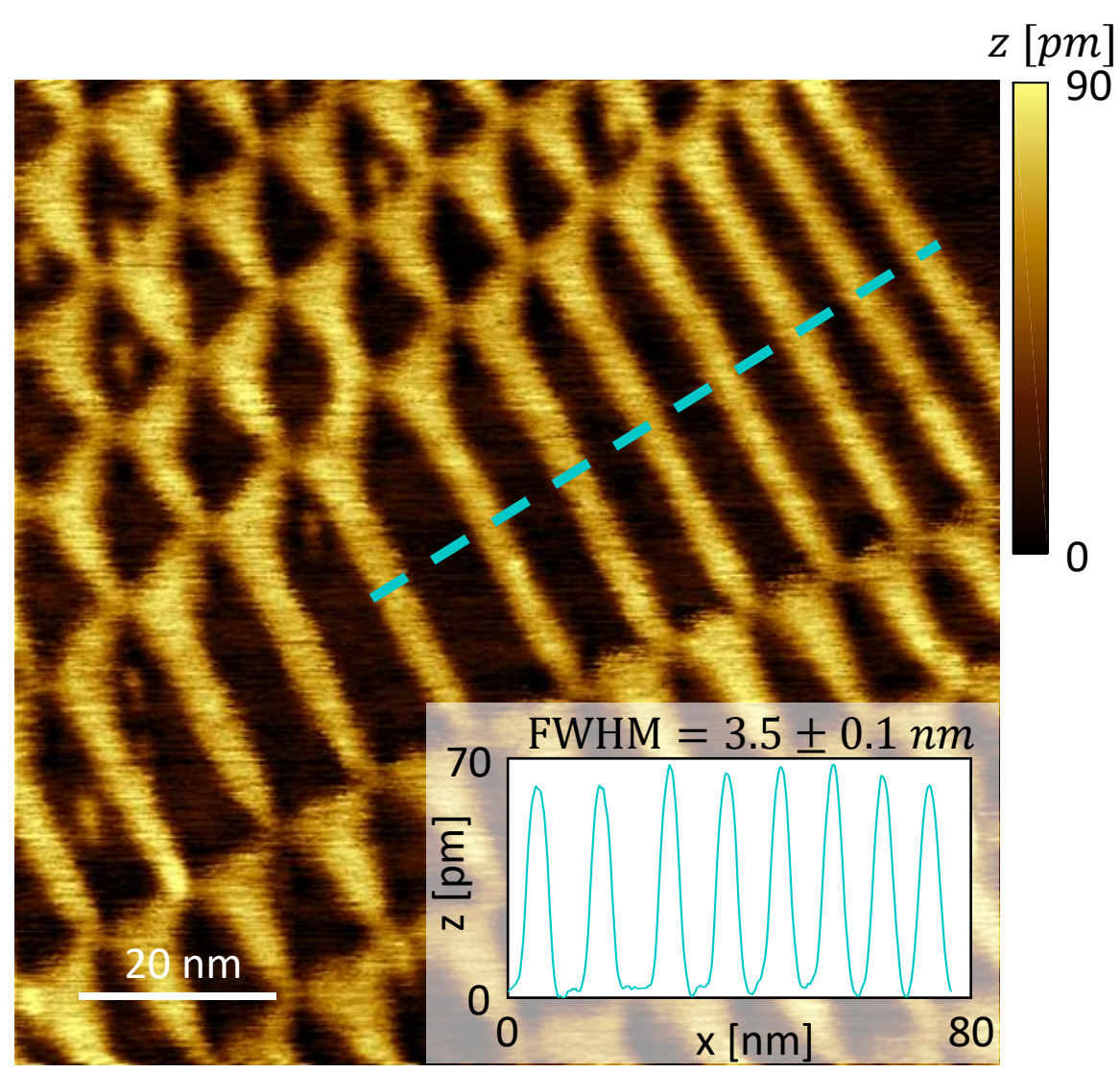

**Figure S2|H-stacked MoSe2/WSe2 DDW width from STM measurements (constant current mode).** Inset: a cross-section of along the dashed marked line, showing a full-width-half-maximum (FWHM) DDW width of $3.5 \pm 0.1$ nm.

S3 presents another STM scan (constant current mode) of a similar sample, showing a series of parallel formation of DDWs. The line-cut (inset) reveals a full-width-half-maximum DDW width of 3.5 nm. A 1D relaxation calculation was performed using the calculated parameters for $MoSe_2/WSe_2$ of Table 1 to resolve the calculated DDW structure. In order for the DDW to be a well-defined 1D structure, the mismatch had to be artificially removed by equating the lattice constant of the two materials (in the 2D calculation, no such simplification was made). The calculated DDW width using the parameters of Table 1 is 3.4 nm, with excellent agreement with the experimental observation.

## S 4. Study of TBG generalized stacking fault energy function using single vs. double domain wall formation

In this section we provide a detailed account of the twisted bilayer graphene (TBG) analysis presented in the main text. The analysis in this section is based on the non-local photocurrent image of Fig. S3a. The non-local photocurrent technique is a novel imaging technique that differ from the more conventional thermo-electric based nano-photocurrent imaging by the origin of the photocurrent signal. In the conventional thermo-electric case, photocurrent is generated through the Seebeck effect as a convolution between the hot spot generated by the enhanced electric field under the tip, and local variations in the Seebeck coefficient. In our case due to the low temperature and high doping the generated temperature profile is broadened (Fig. S3b – top left), and the local photocurrent contribution is suppressed. However, due to the existence in our case of an interface between monolayer (MLG) and TBG, which provides a sharp step in Seebeck coefficient (Fig. S3b – top right), photocurrent is actually generated at the interface, away and independent of tip position. As a result, the technique is sensitive to local changes in optical conductivity governing local absorption (illustrated by Fig. S3b – bottom), and reveals in great details all subcomponents of the structure: from AB and BA domains (orange and cyan dots in Fig. 2a), SDWs and DDWs, AA sites (red dot in Fig. S3a) and plasmonic fringes in the MLG.

The single tuning parameter model presented in the main text attempts to capture the essence of the process governing the formation of the network of domain walls as a competition between SDWs and DDWs. It is an oversimplification in the sense that it neglects the angular dependence of the energy cost per unit length of the domain walls, but it is still useful for the discussion. This model reduces the competition between SDWs and DDWs to one ratio, $\bar{\beta}$, of the energy per unit length of a DDW oriented along the shear direction of the DDW dislocation, and the energy per unit length of a SDW oriented along the same direction (of minimal energy for forming DDW). The rationale is that in order to form a DDW, two SDWs have to be brought together, and bend to reorient along the same direction. This quantity captures the effective attraction between two proximate SDWs, and therefore plays an important role.

Next we anchor the vertices of a given triangle at the AA sites (red dot in Fig. S3a) and minimize the total energy of a triangle by allowing two SDW to collapse and form a segment of a DDW. Fig. S3c schematically shows how different triangular arrangements of the AA sites would minimize the energy as a function $\bar{\beta}$. In the trivial case at $\bar{\beta}$=2 (Fig. S3c top raw, and similarly the fit of Fig. S3d), there is no benefit in forming DDWs, and the minimal energy would yield triangular domains. In the low $\bar{\beta}$ limit, in particular $\bar{\beta} = \sqrt{3}$ (bottom raw in Fig. S3c), all SDW for every triangular geometry would collapse to DDW intersecting at the Fermat point of the triangle (Fig. S3f). We can use $\bar{\beta}$ as a fitting parameter and search for the value that would reproduce the experimental network of Fig. 2a (the full span of values is presented in Supplementary Video 1). The predicted network is strongly dependent on the fitting parameter, and the optimal fit is achieved for $\bar{\beta} = 1.90$ (Fig. S3e) with surprisingly good agreement with experimental data, considering the oversimplification of these assumptions.

To quantify moiré networks constraints on the stacking energy landscape, we spanned according to the value of $\bar{\beta}$ all possible GSFE's for TBG (satisfying the symmetry of TBG) using a 2D dimensionless parameter space $(\zeta, \tau)$ (as schematically illustrated in Fig. 2b). $\zeta$ is the ratio of the GSFE value at the saddle-point (SP) to the value at the AA configuration, and $\tau$ controls flatness of the SP as the second derivative along the line connecting two AA sites (normalized by the GSFE value at the AA configuration). For each point on the $(\zeta, \tau)$ plane, corresponding to one GSFE candidate, we solve a set of 1D relaxation problems describing the profiles of SDW and DDW at different domain wall orientations (see SI S1 for more details on relaxation codes), and extract $\bar{\beta}$. The result is summarized in Fig. S3g, and defines a region (green band) in parameter space that would meet the experimental $\bar{\beta} = 1.90$ criterion.

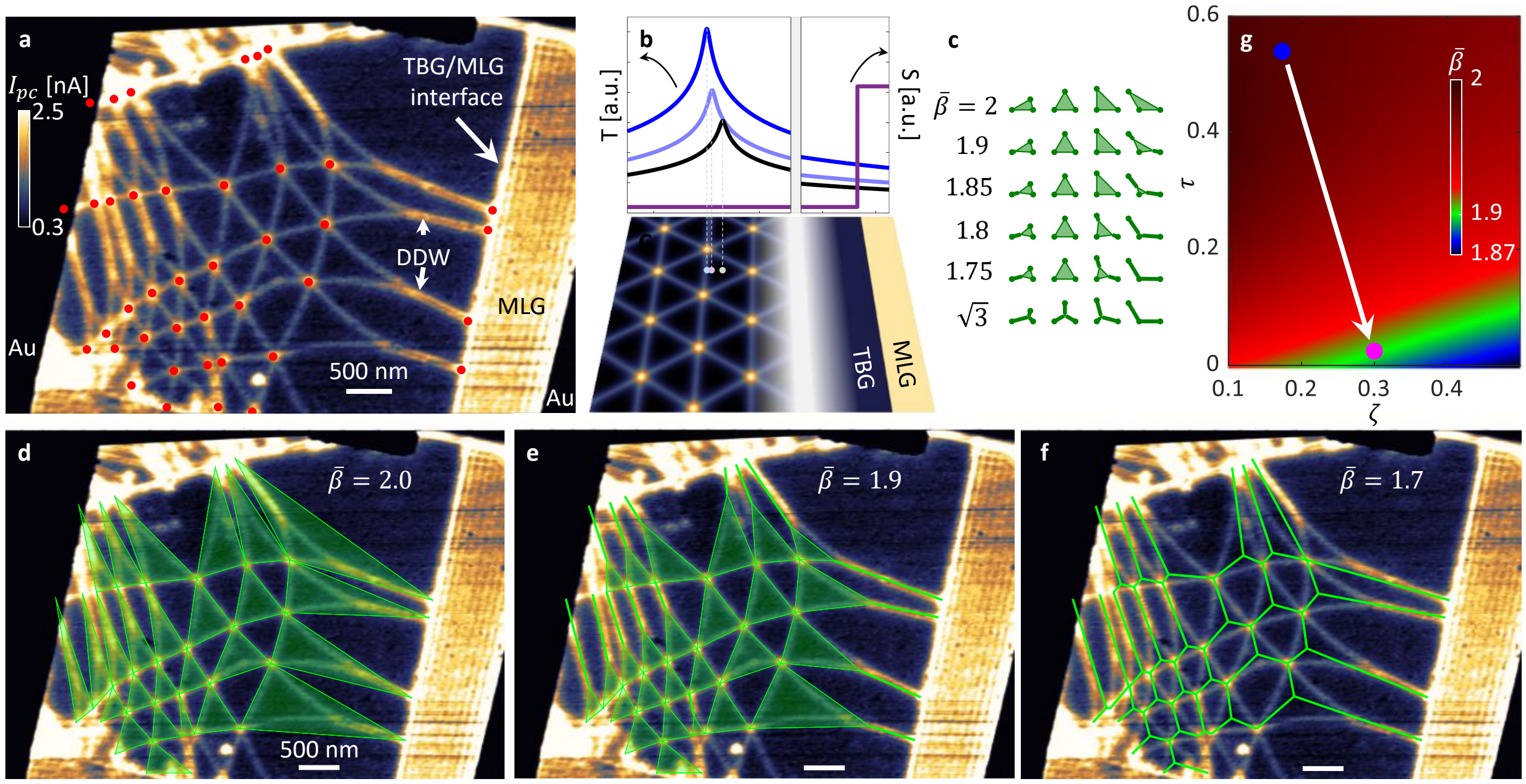

**Figure S3 | Single tuning parameter model for domain formation in TBG. a,** Non-local photocurrent map of moiré super-lattice of a TBG sample. Gold contact (marked by Au) on the left is grounded and the one on the right collects current through a trans-impedance current preamplifier. The map is the collected current for a given tip position (see methods for more information). Temperature is 100 K and sample carrier concentration is tuned to $3 \cdot 10^{12}\ \mathrm{cm}^{-2}$. Data reveal the formation of double domain wall (an example is marked by "DDW") separating domains of similar stacking configurations. **b,** Schematic description of the non-local photocurrent imaging technique. The measurement regime leads to a large thermal cooling length (relative to system size) resulting in delocalized temperature profiles (demonstrated by top- panel representing temperature profiles for different tip positions). In addition, the MLG/TBG interface acts as a Seebeck coefficient step (right top panel) and dominates photo-current generation. Non-local photocurrent should be regarded as a measure of local optical conductivity. **c,** Single fitting parameter model ($\bar{\beta}$ – see text for details) describing domain formation as a competition between single (SDW) and double (DDW) domain walls. In each domain the AA sites (triangle vertices) are fixed and SDW (thin lines) can collapse to form DDW (thick lines) to minimize total energy. This panel shows how different triangles (columns) would collapse for selected $\bar{\beta}$ values (rows). **d-f,** Domain formation within the model parameter model for different fitting parameter values (as mentioned) and AA sites position (marked by red dots in **a**), overlaid on top of experimental data. **g,** Mapping of candidates for GSFE of TBG according to the extracted value of $\bar{\beta}$ (calculated by 1D relaxation calculations – see SI S1 for more details). Curves of GSFE candidates in Fig. 2b are marked by similar colors here.

In the main text we explored the effect of shifting the GSFE parameters toward that region on DDW formations, and observed by 2D relaxation calculations that the experimental observation is reproduced once the TBG GSFE is adjusted as to satisfy $\bar{\beta} = 1.9$ (which is not the case for the existing GSFE parameters in the literature). For completeness, and for practical comparisons with future modelling, we provide the full dependence of the different extracted properties of domain walls on TBG GSFE parameters. Here we assume the mechanical parameters (K and G) remain fixed and follow the literature values mentioned in Table 1. Under these conditions we provide in Table 2 expressions for the domain walls parameters (normalized by $\alpha E_{AA}$, where $E_{AA}$ is the stacking energy for the AA configuration, and $\alpha$ is the atomic u.c lattice constant). Each parameter is fitted with a third order polynomial in $\zeta, \tau$, and we provide $b_{l,m}$ as the coefficient of the $\zeta^l \tau^m$ term.

| DDW parameter | $E_{S,SDW}$ $/\alpha E_{AA}$ | $E_{T,SDW}$ $/\alpha E_{AA}$ | $E_{S,DDW}$ $/\alpha E_{AA}$ | $E_{T,DDW}$ $/\alpha E_{AA}$ | $\bar{\beta}$ |
|---|---|---|---|---|---|
| $b_{0,0}$ | 2.64 | 4.147 | 5.789 | 7.519 | 1.9480 |
| $b_{0,1}$ | 1.257 | 1.975 | 5.628 | 4.250 | 0.2944 |
| $b_{0,2}$ | -0.127 | -0.200 | -5.381 | -1.049 | -0.4872 |
| $b_{0,3}$ | 0.010 | 0.0150 | 2.887 | 0.252 | 0.2469 |
| $b_{1,0}$ | 35.393 | 55.606 | 77.790 | 101.181 | -0.2379 |
| $b_{1,1}$ | -3.005 | -4.720 | -0.453 | -8.603 | 0.2722 |
| $b_{1,2}$ | 0.229 | 0.359 | -2.820 | 1.104 | -0.0856 |
| $b_{2,0}$ | -42.950 | -67.468 | -99.327 | -122.838 | 0.1834 |
| $b_{2,1}$ | 2.850 | 4.477 | 5.834 | 7.522 | -0.1251 |
| $b_{3,0}$ | 28.682 | 45.058 | 66.590 | 82.285 | -0.0791 |

Table 2: TBG candidates that satisfy $\bar{\beta} = 1.90$.

In Fig. 2d-e we use relaxation calculations (see SI section S1 for more details) to compare the commonly accepted GSFE version for TBG in literature[31], labeled "Carr et al." with one representative GSFE candidate on the $\bar{\beta} = 1.9$ band (Fig. S3g), labeled "moiré constrained GSFE". Here we provide further details about these calculations. Figure S4 shows the stacking energy density for the two GSFE versions (literature and moiré constrained GSFE encircled by blue and magenta rectangles respectively). For each GSFE we explore different initial and boundary conditions (IC and BC) leading to the relaxed lattice solutions (second and fourth columns in Fig. S4). In the first case (Fig. S4a-d), the stacking configuration is forced in a few selected points (dots in a and c), and the IC is spline-interpolated between these points. After relaxation (Fig. S4b and d for literature and moiré constrained GSFE respectively) shows the formation of SDWs and DDWs. Interestingly, the moiré constrained GSFE solution shows a symmetric DDW formation reflecting the AB/BA domain wall symmetry. While this is expected, it does not agree with the experimental observation (Fig. 2). Even using an IC which breaks the AB/BA symmetry, by taking the single tuning parameter network as the input into the relaxation calculation (Fig. S4e-h) leads to a similar solution (compare Fig. S4b-f and d-h). Despite the obscure experimental origin of this symmetry breaking, we can still account for it by adding BC points that break the symmetry, as shown in Fig. S4i,k leading to the solutions of Fig. 2d-e (Fig. S4i and S4l respectively). Note that the solutions have similar total energies pointing to the degenerate nature of the formed domain wall network.

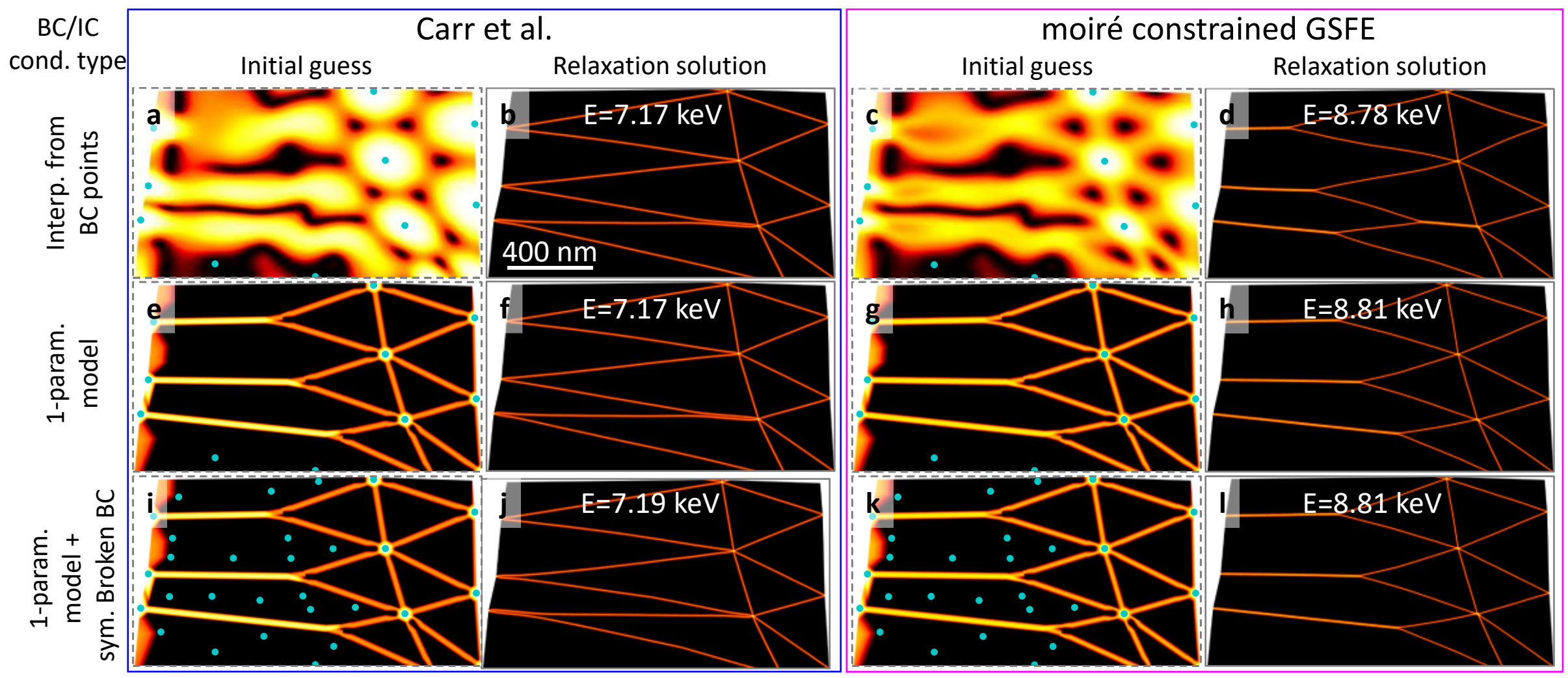

**Figure S4 | Comparing GSFE candidates of TBG through relaxation calculation. a-b,** Energy density map of initial guess (**a**) used for the relaxation calculation result (**b**) using the GSFE of Carr et al.[31]. The dots mark points of fixed configuration boundary conditions (BC). The map of **a** was interpolated from the stacking configurations at the dots. **c-d,** Similar to **a-b** but for the moiré constrained GSFE candidate discussed in Fig. 2 and in this section. **e-h,** Similar to **a-d** but with initial conditions set by the single tuning-parameter model with $\bar{\beta}$ (see section text). **i-l,** Similar to **e-h** but with additional BC dots breaking the symmetry between AB and BA domains. All panels share the scale-bar of **b**. Panels **a-b,e-f,i-j** and **c-d,g-h,k-l** share the color-maps of Fig. 2d and Fig. 2e respectively. Total calculated energies after the relaxation process are indicated on each relaxation calculation result, showing the near degenerate nature of the solutions.

## S 5. Defect-induced doping in TDBG stacks

Bilayer graphene flakes that were used to make TDBG stacks for this work were cut in two ways: the 'tear and stack' approach, and by anodic-oxidation lithography (see sample preparation section under Materials and methods). Some of the stacks were measured while still on a transfer slide (stacked on top of a PPC/PDMS stamp/glass slide structure), while others were made into full devices and measured at controlled electrostatic environment. In the histogram of Fig. 3a only data of well-defined electrostatics were used (for the full set of data, with an additional large $\kappa^{-1}$ valued domains, see SI section S9 below). In a few of the anodic-oxidation cut stacks, and most pronouncedly in the case presented in Fig. 3e, the stacks seem to have a high level of defect-induced doping. This is apparent from both the enhanced ABAB/ABCA nearfield contrast, as well as by the flattening of the SDW's (as discussed in Fig. 3 and relevant text). The origin of this defect-induced doping is not known, but it is hypothesized to be related to full/partial oxidation of the layer that was cut during the anodic oxidation, perhaps in cases of extremely high humidity levels during the cutting process. Starting from the DFT-D2 approach for TDBG, we examined the evolution of the energy per unit area of the rhombohedral domain, $\sigma$, as a function of doping and interlayer bias, based on DFT calculations (as described in Materials and methods). The results are presented in Table 3 below, and they reveal a relatively weak dependence of the curvature on the interlayer bias and n-doping, but a considerable effect of p-doping.

Assuming the change of the curvature in the case of Fig. 3e is mostly from doping, we estimate the doping level from curvature to be roughly $8 \cdot 10^{12}\ cm^{-2}$. At that doping level the GSFE function is defined (as calculated by DFT methods) by the following coefficients: $c_0 = 10.2226$, $c_1 = 6.0451$, $c_2 = -0.4919$, $c_3 = -0.2199$, $c_4 = -0.0037$, $c_5 = -0.0073$ (all in units of $meV/u.c.$). These are the parameters used in Fig. 3b (dashed green) and green dots of Fig. 3f.

| $n\ [10^{12}\ cm^{-2}]$ | -9.5 | -5.7 | 0 | 5.7 | 9.5 |
|---|---|---|---|---|---|
| $\sigma\ [meV/nm^2]$ | 3.934 | 4.445 | 4.637 | 1.366 | -0.583 |

| $D\ [V/nm]$ | 0 | 0.2 | 0.6 | 1 |
|---|---|---|---|---|
| $\sigma\ [meV/nm^2]$ | 4.637 | 4.499 | 3.629 | 2.072 |

Table 3: $\sigma$ in as a function of interlayer bias and doping.

## S 6. Examples of images for TDBG domains curvature extraction

Fig. S5 presents 3 additional representative examples of nearfield (Fig. S5a), STM (Fig. S5b) and STS (Fig. S5c) imaging of TDBG moiré superlattice that were used in constructing the histogram of Fig. 3a. The histogram is based on a total of 31 scans, most of which are not explicitly shown here, but will be provided upon reasonable request.

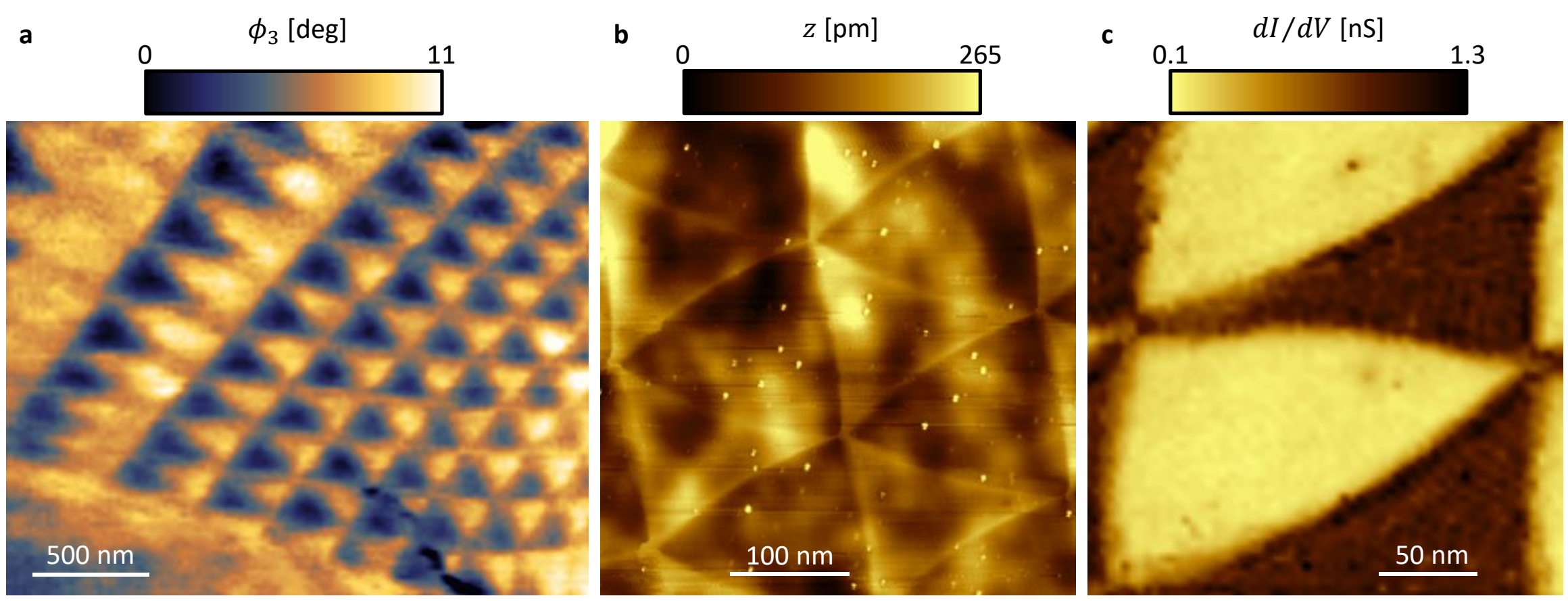

**Figure S5|Examples of images used for curvature extraction of TDBG domains. a,** Mid-IR (940 $\mathrm{cm}^{-1}$) nearfield phase (ABCA – dark, ABAB - bright) similar to Fig. 1c. **b**, STM scan showing ABAB (dark) and ABCA (bright) domains. **c**, STS imaging (ABCA – dark, ABAB – bright) similar to Fig. 1b.

## S 7. The 2D soap-bubble model of TDBG

The 2D soap-bubble aims to describe the shape of rhombohedral (ABCA) domains surrounded in a Bernal background (ABAB) as a competition between single domain walls (SDWs), double domain walls (DDWs) and domain area. A preliminary version, restricted to the SDW interface has been presented recently[38]. Here we generalize it to include DDWs, which dominate the structure at the low twist angle limit as well as under external strain. The results of this model were used in the main text (dashed green lines in Fig. 3c-d and solid lines in Fig. 3e), with great agreement with full relaxation calculations except for a narrow angular region in Fig. 3e where DDWs start to form.

We start by describing the interface between ABCA and ABAB in absence of the formation of DDWs. We assume the energy of such an interface is composed of an energy per unit area $\sigma$ (which is directly extracted from the difference in GSFE values at ABCA and ABAB), and an energy per unit length of a SDW. The latter has an angular dependence of the form $\gamma_1(\varphi) = \left(E^n_{S,SDW}\cos^2\varphi + E^n_{T,SDW}\sin^2\varphi\right)^{\frac{1}{n}}$. First, we assume that the domain wall shape is described by a function $y = f(x)$ with boundary conditions $f(x=0) = f(x=L) = 0$. We can then write an expression for the energy of the domain as: $E = \int_0^L dx\mathcal{E}(x,f,f')$, where the energy functional is $\mathcal{E}(x,f,f') = \sqrt{1+f'^2}\gamma_1\big(\varphi(f')\big) + \sigma f$, and the domain wall orientation angle function $\varphi(f') = \tan^{-1} f'$ . After writing Euler-Lagrange equations and substituting $\mathrm{f}' = \tan\varphi$ one can show:

$$x(\varphi) = \frac{1}{\sigma}(\gamma_1(\varphi)\sin\varphi + \gamma_1{}'(\varphi)\cos\varphi) + const$$

Furthermore:

$$y'(\varphi) = x'(\varphi)\tan\varphi$$

$$y''(\varphi) = \frac{1}{\cos^2\varphi}x'(\varphi) + \tan\varphi\, x''(\varphi)$$

And one can show that the curvature satisfies:

$$\kappa(\varphi) \equiv \frac{x'y'' - y'x''}{(x'^2 + y'^2)^{\frac{3}{2}}} = \frac{\sigma}{\gamma_1(\varphi) + \gamma_1{}''(\varphi)}$$

If we approximate $\gamma_1(\varphi)$ to an analytic form of $n = 2$, this would yield an elliptic SDW shape with major (oriented along the shear direction) and minor half axis (A and B respectively) such that:

$$A = \frac{1}{\sigma}E_{T,SDW}, B = \frac{1}{\sigma}E_{S,SDW}$$

For a given GSFE of TDBG within different explored approaches we extract $E_{S,SDW}, E_{T,SDW}$ by solving a 2D relaxation problem such as in Fig. S6a, extract A and B and multiply by $\sigma$. $\sigma$ can be evaluated directly from the GSFE as $\sigma = \frac{1}{S_{uc}}(c_4 + c_5)$, where $S_{uc} = \frac{\sqrt{3}}{2}\alpha^2$ is the atomic scale unit cell area (introduced due to the unit choice in Table 1). Due to the instability of the rhombohedral phase in TDBG, the SDW separating the rhombohedral and Bernal phase is not a stable 1D soliton solution. Instead, it is a 2D structure stabilized by effective pressure, originating from the difference in energy cost of the two phases. As a result, in order to calculate $E_{S,SDW}, E_{T,SDW}$ we have to do 2D relaxation calculations. We define a $L_x$ by $L_y$ rectangular geometry, centered at the origin, and assume a boundary between rhombohedral and Bernal phases. We pin the SDW at two points, $x = 0, y = \pm\frac{1}{2}L_y$, by forcing a saddle point configuration at these points. For any point along the mesh edges far enough from the SDW (50 nm in the results below) we forced a Bernal phase on the right and rhombohedral phase on the left. After relaxation an elliptical arc shaped domain is formed, as shown in Figure S6a, suggesting that $\gamma_1(\varphi)$, the angular dependence of the SDW energy cost per unit length mentioned before can indeed be approximated with an exponent of $n = 2$. The major and minor axis of the ellipse (A and B respectively) were extracted by fitting an ellipse (turquoise dashed line in Fig. S6a) for DFT-D2, and similarly for all other DFT approaches considered in this work. These geometrical parameters were directly connected to the energy per unit length coefficients as by the above expressions. One should note, though, that this solution is valid as long as DDWs do not start to form, and that once it becomes energetically beneficial to introduce DDWs (due to strain, or do to geometrical limitations) the pure elliptic solution is no longer valid. The angular dependence of the energy cost of a DDW was extracted by 1D calculation as described in SI section S1, and summarized in Table 1, thus extracting all required parameters for the 2D soap-bubble model.

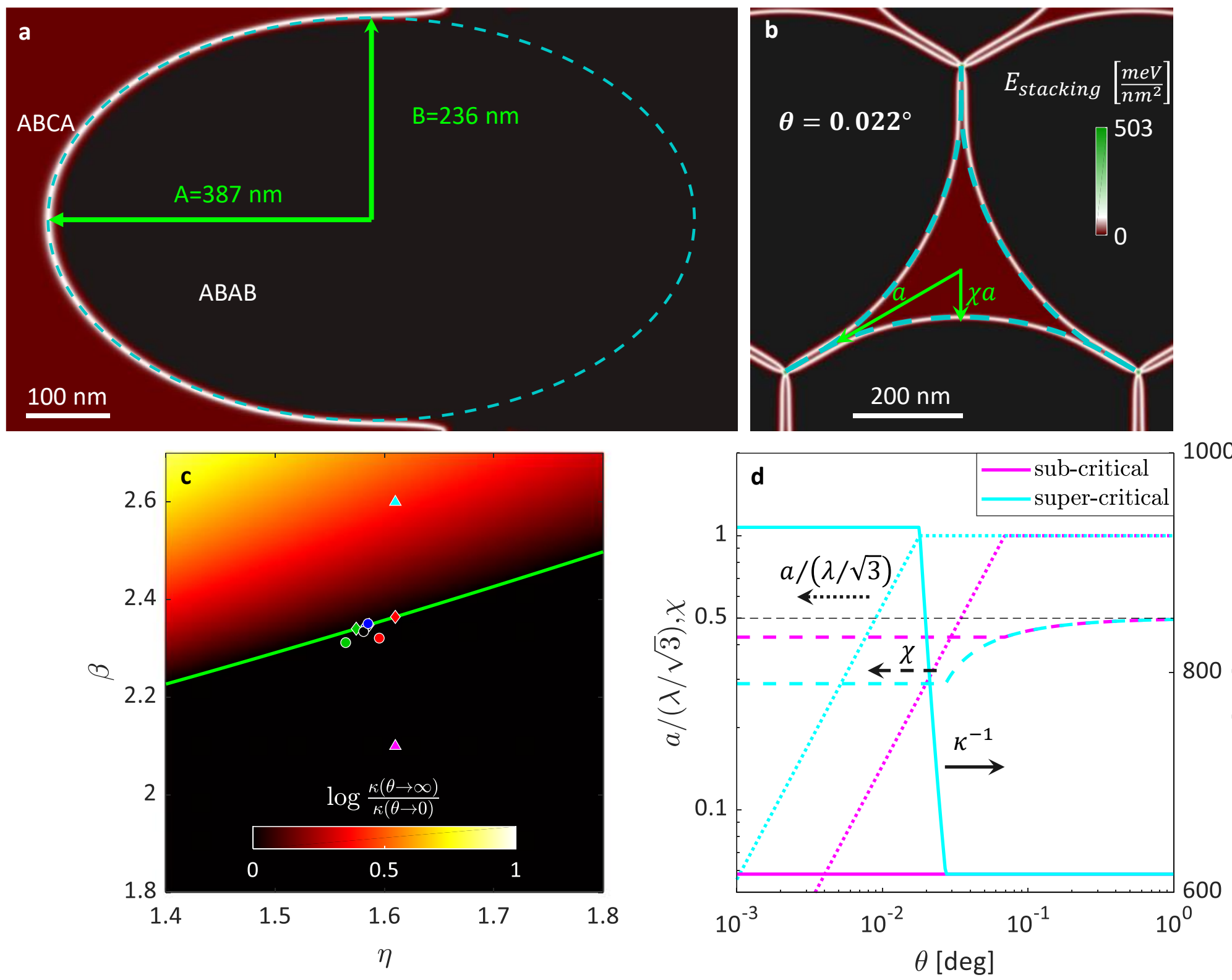

**Figure S6|Analysis of the 2D soap-bubble model. a,** Stacking energy density from a 2D relaxation calculation of the interface between ABCA (left of interface) and ABAB (right of interface) stackings for the GSFE of DFT-D2, yielding an elliptic-arc shaped domain wall. The extracted major (A) and minor (B) axis of the formed ellipse (fit in dashed green line) are directly used as inputs to the 2D soap-bubble model. White double arrow marks shear direction for the SDW dislocation. **b,** Solution of a continuous relaxation model (false color: stacking energy density) and soap-bubble model domain shape (dashed green) for a twist angle of 0.022° for DFT-D2. Geometrical parameters of the soap-bubble model are illustrated. **c,** Phase space revealing a phase transition between a single valued curvature below (the sub-critical regime) a critical line for $\beta_c(\eta) = \sqrt{\eta^2 + 3}$ (green line), and a double asymptote behavior above the critical line (the super-critical regime). The diagram plots the log of SDW curvature asymptotes ratio (at infinite and 0 twist angles) as a function of two dimensionless energy ratios defining the phase-space ($\beta = \gamma_2/E_{S,SDW}$ and $\eta = E_{T,SDW}/E_{S,SDW}$).The markers show the location of the 4 DFT approaches used in this work (with similar color convention as in Fig. 3), plus two artificially added examples (triangles) to demonstrate sub-critical (magenta) and super-critical (cyan) behaviors. Circles are the locations in phase space of the different DFT approaches as extracted directly, and diamonds are fits forced to be along the critical line (see text for more information). **d,** Analysis of the domain geometrical parameters $(a, \chi)$ (left y-axis) and resulting inverse curvature (right y-axis) for the sub-critical (magenta) and super-critical (cyan) examples marked in **c**, showing the qualitative difference between the two regimes in the soap-bubble model.

For a given position of the domain vertices (the BAAC stacking locations), the 2D soap-bubble model thus assumes the energy has three contributions: an area term, an energy per unit length term of the SDW and an energy per unit length of the DDW. In the simplest case we treat an equilateral triangular geometry. This will allow us to reach analytic expressions for the solution minimization the energy, and explore the energy functional for the meta-stability discussion as presented in SI section S9. Under these assumptions the DDW is aligned along the shear direction, and therefore the angular dependence of $\gamma_2(\varphi)$ is reduced to $\gamma_2 = \gamma_2(\varphi = 0)$. We further assume that the SDW segments have an elliptic form with major to minor axis ratio of $\eta$. We assume the ellipse is oriented along the shear direction of the SDW, meaning in our simplified geometry, along the lines connecting the vertices, but relax the ellipse dimensions by allowing the minor axis dimension to vary as to minimize the energy. In the case of no DDWs the energy will be minimized by a choice of A, B as dictated by the above Euler-Lagrange equation solution. However, once DDWs start to form that will no longer necessarily be the case. We define two geometrical parameters controlling the domain shape $a$ and $\chi$ such that $a$ is the distance from the triangle center to the end point of one of the DDWs and $\chi a$ is the distance from triangle center to nearest point on the SDW (see notation on representative domain of Fig. S6b). $\lambda$ is the moiré length and the edge of the triangle. Within this notation the shape of one of the 3 domain wall curves defining the domain would be:

$$y(x) = \begin{cases} \dfrac{x}{\sqrt{3}} & 0 \le x \le \dfrac{\lambda}{2} - \dfrac{\sqrt{3}}{2}a \\ y_0 + \sqrt{B(a,\chi,\eta)^2 - \eta^{-2}\left(x - \dfrac{\lambda}{2}\right)^2} & \left|x - \dfrac{\lambda}{2}\right| \le \dfrac{\sqrt{3}}{2}a \\ \dfrac{\lambda - x}{\sqrt{3}} & \dfrac{\lambda}{2} + \dfrac{\sqrt{3}}{2}a \le \lambda \end{cases}$$

Where:

$$B(a,\chi,\eta) = \frac{\dfrac{3}{4\eta^2} + \dfrac{1}{4}(1-2\chi)^2}{1-2\chi}a$$

$$y_0(a,\chi,\eta) = \frac{\lambda}{\sqrt{12}} - \chi a - B(a,\chi,\eta)$$

And the parameters are defined within the boundaries:

$$0 < a \le \frac{\lambda}{\sqrt{3}}; \zeta_\eta \le \chi < \frac{1}{2};\ \zeta_\eta = \frac{\sqrt{\eta^2+3}}{2\eta^2}\left(\sqrt{\eta^2+3} - \sqrt{3}\right)$$

Where $\zeta_\eta$, the bottom limit on $\chi$, is set such that the SDW elliptic segment is tangent to the DDW. Under such assumptions, the total energy is the sum of the SDW and DDW contributions and the domain contributions, $E = E_1(a,\chi,\eta) + E_2(a) + E_\Sigma(a,\chi,\eta)$. Where:

$$E_1(a,\chi,\eta) = 6\eta E_{S,SDW} B(a,\chi,\eta) \int_{\frac{\pi}{2}-\Phi(a,\chi,\eta)}^{\frac{\pi}{2}} d\phi \frac{\sqrt{\eta^4 \sin^2\phi + \left(\frac{E_{T,SDW}}{E_{S,SDW}}\cos\phi\right)^2}}{(1+(\eta^2-1)\sin^2\phi)^{\frac{3}{2}}}$$

$$E_2(a) = 3\gamma_2\left(\frac{\lambda}{\sqrt{3}} - a\right)$$

$$E_\Sigma(a,\chi,\eta) = 3\sigma\left(\sqrt{\frac{3}{4}(1-\eta^{-2})a^2 + B(a,\chi,\eta)^2}\big(\chi a + B(a,\chi,\eta)\big)\sin\Phi(a,\chi,\eta) - \eta B(a,\chi,\eta)^2\left(\frac{\pi}{2} - \tan^{-1}(\eta\cot\Phi(a,\chi,\eta))\right)\right)$$

$$\sin\Phi(a,\chi,\eta) = \sqrt{\frac{\frac{3}{4}a^2}{\frac{3}{4}(1-\eta^{-2})a^2 + B(a,\chi,\eta)^2}}$$

If we further assume $\eta = \frac{E_{T,SDW}}{E_{S,SDW}}$, as would be the case for the energy minimizing shape of a SDW in the absence of formation of DDWs, then the expression for $E_1(a,\chi,\eta)$ can be simplified to the analytic form:

$$E_1(a,\chi,\eta) = 6\eta E_{S,SDW} B(a,\chi,\eta)\left(\frac{\pi}{2} - \tan^{-1}(\eta\cot\Phi(a,\chi,\eta))\right)$$

This defines a two dimensional phase-space for the domain energy, as presented in the Fig. S6c and specific examples of sub-critical and super-critical twist angle dependence presented in Fig. S6d. Mapping the phase-space, one can show that there is a second order phase transition as a function of $\beta \equiv \frac{\gamma_2}{E_{S,SDW}}$, across $\beta_c(\eta) = \sqrt{\eta^2+3}$. Below this line, the sub-critical regime (represented here by the magenta triangle in Fig. S6c), there is a single valued curvature of the SDW (as a function of twist angle) satisfying $\kappa_\infty^{-1} = \eta^2 \frac{E_{S,SDW}}{\sigma}$. Above the line, the super-critical regime (represented here by the cyan triangle in Fig. S6c), the curvature covers a span of values between two asymptotic values. For a large twist angle the system converges to a similar domain structure as in the sub-critical regime, but at the low twist angle limit, a different solution emerges satisfying the ratio: $\frac{\kappa_{\theta\to 0}^{-1}}{\kappa_{\theta\to\infty}^{-1}} = 1 + \frac{1}{\beta_c}\frac{\beta-\beta_c}{1-\frac{\sqrt{3}}{\eta}\left(\frac{\pi}{2}-\tan^{-1}\frac{\sqrt{3}}{\eta}\right)}$. The two regime also differ in terms of the path the energy minimizing point takes on the $(a,\chi)$ plane as a function of twist angle, as shown in Fig. S6d. In the super-critical regime (cyan case in Fig. S6d), the energy reaches its minimal value on the edges of the phase-space. The super-critical of Fig. S6d reveal three regimes as a function of twist angle. The first regime, of SDW deformation, is revealed for large twist angles, while $\lambda < \lambda_{c2} = \frac{2}{\beta_c}\kappa_{\theta\to\infty}^{-1}$. In that regime the energy minimizing point travels along the right edge of the $(a,\chi)$ plane at $(a,\chi) = \left(\frac{\lambda}{\sqrt{3}}, \frac{1}{2} + \frac{\sqrt{3}}{2\eta}\left[\sqrt{\left(\frac{\beta_c\lambda_{c2}}{\eta\lambda}\right)^2 - 1} - \frac{\beta_c\lambda_{c2}}{\eta\lambda}\right]\right)$. As $\lambda$ approaches $\lambda_{c2}$ , $\chi$ approaches $\zeta_\eta$, the bottom edge of the $(a,\chi)$ plane. For intermediate values of $\lambda_{c2} \le \lambda \le \lambda_{c1} = \frac{2}{\beta_c}\kappa_{\theta\to 0}^{-1}$ the energy will be minimized for a fixed point of $(a, \chi) = \left(\frac{\lambda}{\sqrt{3}}, \zeta_\eta\right)$. In this narrow angular regime, the 'self-similar regime', the domain expands self-similarly as the twist angle decreases. As the twist angle further decreases, $(a, \chi)$ do not change with $\lambda$, staying fixed at $(a, \chi) = \left(\frac{\lambda}{\sqrt{3}}, \zeta_\eta\right)$. In this regime, of DDW formation, the rhombohedral domain no longer changes and remains fixed while DDWs extend from its vertices, effectively traveling to the left along the bottom edge of the $(a, \chi)$ plane.

In contrast, in the sub-critical regime (magenta line in Fig. S6d), when $\beta \leq \beta_c$, the SDW deformation stage ends before reaching the bottom edge of the $(a, \chi)$ plane. Throughout this entire regime $\chi$ satisfies: $\chi(a) = \frac{1}{2} + \sqrt{\left(\frac{\kappa^{-1}_{\theta\to\infty}}{\eta^2 a}\right)^2 - \frac{3}{4\eta^2}} - \frac{\kappa^{-1}_{\theta\to\infty}}{\eta^2 a}$. Once $\lambda$ reaches $\sqrt{3}a_c$ where $a_c = \frac{2}{\beta_c^2}\left(\frac{\beta}{\sqrt{3}} - \frac{\sqrt{\beta_c^2 - \beta^2}}{\eta}\right)\kappa^{-1}_{\theta\to\infty}$, DDW start to form, the energy will be minimized for a fixed point of $(a, \chi) = \left(a_c, \chi(a_c)\right)$ and the rhombohedral domain will no longer change with decreasing twist angle.

It is important to note that the soap-bubble model agrees very well with the full relaxation solution, as shown in the main-text and Supplementary Videos 2-5. However, for the super-critical case, there is a narrow angular range between the 'SDW deformation' regime to the 'DDW formation' regime, in which the two models disagree. The full relaxation solution does not show a clear self-similar expansion, as predicted by the soap-bubble model in the super-critical regime. This is probably due to the simplistic representation of the DDW in the soap-bubble model, that doesn't consider a realistic gradual formation of a DDW by collapsing two single SDWs.

Surprisingly, all 4 DFT approaches considered in this work for describing TDBG fall extremely close to the critical line (see circle in Fig. S6c relative to green line), with $\beta$ within about 1% of $\beta_c$. Therefore, and due to higher numerical precision of the calculation of $E_{s,DDW}$ (from 1D models) and $E_{s,SDW}$ (from moiré domain relaxation calculations as in Fig. 3c-d) compared to extracting the elliptic arc shape of a SDW (as in Fig. S6a), in our modelling of TDBG in this work we assumed all 4-models (with parameters listed in Table 1) fall on the critical line (diamonds in Fig. S6c) but forcing the extracted $E_{s,DDW}$ and $E_{s,SDW}$.

## S 8. Moiré domains deformation under strain

As shown experimentally, moiré patterns in the low twist angle limit are highly susceptible to strain, resulting in some cases in elongated 1D structures and formation of double domain walls. In this section we will present an analytical description of the origin of this behavior.

We restrict our discussion to the case of a fixed external strain, which in its most general form is assumed to be a 2D symmetric tensor: $\overleftrightarrow{\epsilon} = \begin{pmatrix} \epsilon_{xx} & \epsilon_{xy} \\ \epsilon_{xy} & \epsilon_{yy} \end{pmatrix}$. The strain can be written in the following way:

$$\overleftrightarrow{\epsilon} = \epsilon_c I + \epsilon_s(\cos\phi_s\,\sigma_x + \sin\phi_s\,\sigma_z)$$

Where $\epsilon_c = \frac{\epsilon_{xx}+\epsilon_{yy}}{2}$ is a pure dilation strain and $\epsilon_s = \sqrt{\left(\frac{\epsilon_{xx}-\epsilon_{yy}}{2}\right)^2 + \epsilon_{xy}^2}$ is a shear strain term. In this notation $\phi_s$ encodes information about the direction of the shear strain. $\sigma_x$ and $\sigma_z$ are Pauli's matrices in the standard notation. Before any relaxation takes place the displacement field will take the form (setting without loss of generality the displacement field at the origin to 0):

$$\boldsymbol{u}(\boldsymbol{r}) = \epsilon_c \boldsymbol{r} + \epsilon_s(\cos\phi_s\,\sigma_x + \sin\phi_s\,\sigma_z)\boldsymbol{r}$$

When $\epsilon_s$ is sufficiently large a critical behavior emerges, in which for a finite $\epsilon_s = s_c$ the moiré unit cell collapses to form a 1D structure, as indeed observed experimentally. This behavior can be seen from deriving expressions for the moiré unit cell lattice vectors, $V_1, V_2$, which take the form (following the notation of the previous section):

$$V_j = \frac{\alpha}{\Delta}\Big((1-(1-\mu)\epsilon_c)R_{\psi_j+\theta} - (1+\delta)(1+\mu\epsilon_c)R_{\psi_j} + \epsilon_s(1-\mu)R_{\phi_s-\psi_j-\theta} + \epsilon_s\mu(1+\delta)R_{\phi_s-\psi_j}\Big)\begin{pmatrix}1\\0\end{pmatrix};\, j = 1,2$$

Where $\psi_1 = \theta_0$, $\psi_2 = \theta_0 + \frac{\pi}{3}$, $R_\psi = \begin{pmatrix}\cos\psi & -\sin\psi\\ \sin\psi & \cos\psi\end{pmatrix}$ is the rotation matrix by an angle $\psi$, and:

$$\Delta = (\epsilon_c(1-\mu)-1)^2 - (1-\mu)^2{\epsilon_s}^2 + (1+\delta)^2((1+\epsilon_c\mu)^2 - {\epsilon_s}^2\mu^2) - 2(1+\delta)\big(1+(2\mu-1)\epsilon_c - (1-\mu)\mu({\epsilon_c}^2 - {\epsilon_s}^2)\big)\cos\theta$$

We further define the two emerging periods of the moiré super-lattice, $\lambda_j \equiv |V_j|$. From this expression, it is clear that as $\Delta \to 0$, the moiré unit cell would diverge. We define the quantity $s_c$, such that $\epsilon_s = s_c$ leads to $\Delta = 0$. $s_c$ should be regarded as a critical shear strain. Let's examine the area of the moiré unit cell as a figure of merit of this critical behavior. After some further derivation, one can show that it takes the form:

$$S_{moiré} \equiv |V_1 \times V_2| = \frac{\sqrt{3}}{2}\alpha^2 \frac{(s_c^2 - {\epsilon_s}^2)^{-1}}{(1-\mu)^2 + (1+\delta)^2\mu^2 + 2(1+\delta)(1-\mu)\mu\cos\theta}$$

Figure S7 explores the structural dependence of the moiré super-lattice. Fig. S7a-c provide a specific example of the collapse of a unit cell as $\epsilon_s$ approaches $s_c$ for pure shear of a mismatched system (using $MoSe_2$/$WSe_2$ parameters as a representative of such a system). Fig. S7a explores the moiré cell collapse for a strain oriented along the x-axis, Fig. S7b presents the unit cell area as a function of $s_{shear}$ and strain angle (however, curves of all angles collapse). Note that while neither $S_{moiré}$, $\Delta$ nor $s_c$ depend on strain direction (from the above expressions), the exact nature of the collapse as well as relaxation calculations (Fig. S7d-f) are affected by these details.

In order to get some intuition about the expression let's simplify matters by examining a system with no mismatch (therefore $\delta = 0$), such as a homo-bilayer structure. If we further assume for simplicity that one layer is rigid, i.e. $\mu = 0$, we get:

$$s_c^2 = {\epsilon_c}^2 + 4(1-\epsilon_c)\sin^2\frac{\theta}{2};\; S_{moiré} = \frac{\sqrt{3}}{2}\alpha^2(s_c^2 - {\epsilon_s}^2)^{-1}$$

From this expression it is clear that the critical behavior will be observed only if $\epsilon_s$ approaches $\sqrt{{\epsilon_c}^2+4(1-\epsilon_c)\sin^2\frac{\theta}{2}}$, and therefore the dilation strain has a stabilizing effect on the unit-cell, preventing its collapse.

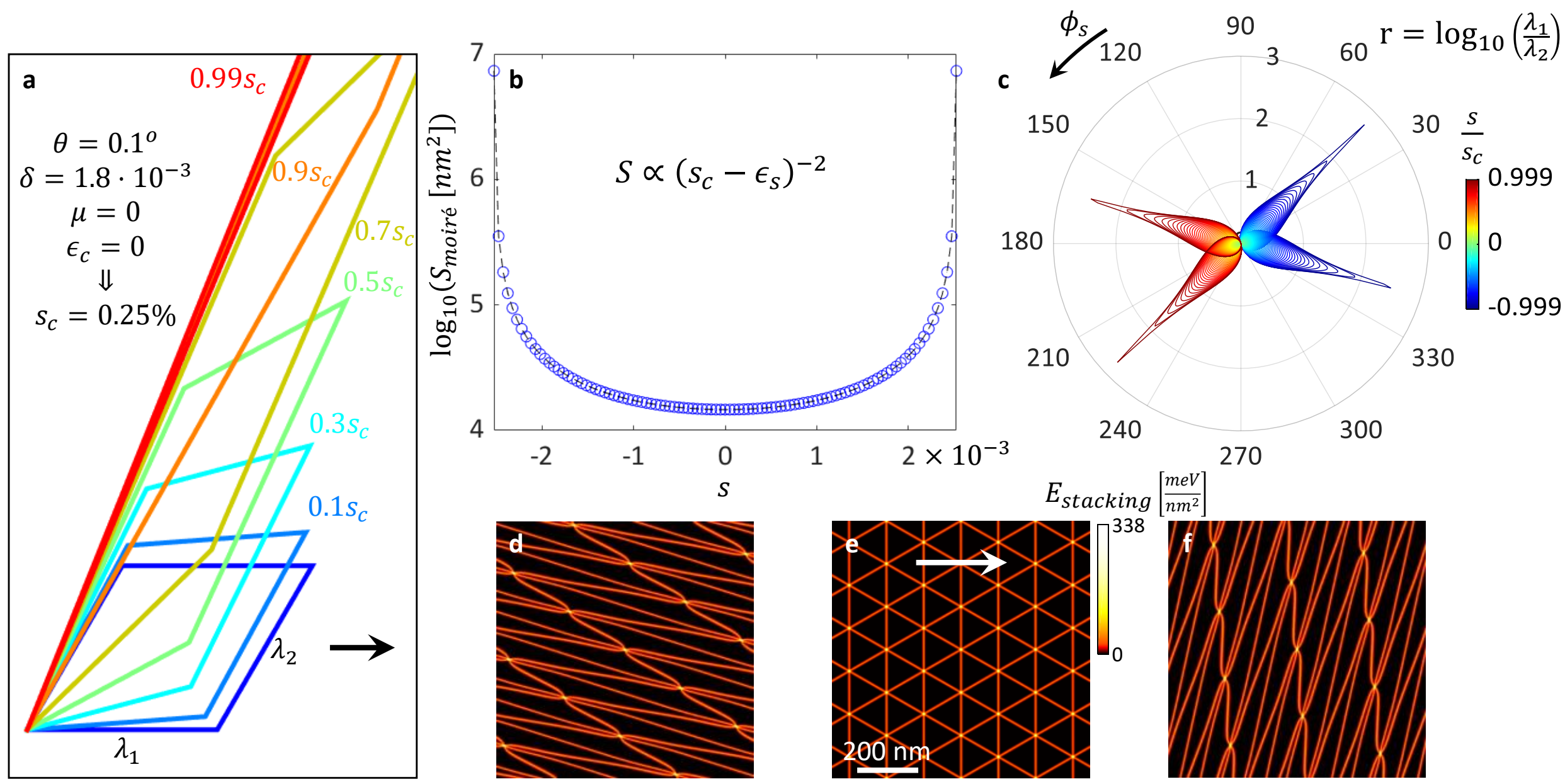

**Figure S7 | Critical structural dependence of the moiré super-lattice on pure shear strain. a-c,** Moiré unit cell collapse due to shear strain for mismatched system with $\theta = 0.1^o$, $\delta = 1.8 \cdot 10^{-3}$, $\mu = 0$, $\epsilon_c = 0$ (see text for details), showing the collapse of the unit cell to a 1D structure toward a critical shear strain value of 0.25% (**a** – strain direction of example marked by arrow), yielding a diverging unit cell area independent of strain orientation (**b**). **c,** Polar plot of ratio of two unit-cell lattice spacings (on a log scale). Color indicates the shear strain value relative to the critical strain. The strong directionality of the pattern suggests that the resulting collapsed patterns depend on strain orientation. **d-f,** Stacking energy density from 2D relaxation calculations of uniaxially strained 0.1° TBG (assuming a Poisson ratio of 0.22), for 3 different strain values: $-0.14\%$ (**d**), $0\%$ (**e**), $0.14\%$ (**f**). White arrow in **e**: The strain orientation for **d-f**.

So far this discussion was somewhat hypothetical, since $\epsilon_c$ and $s$ were treated as independent. In practice, we need to do a more careful analysis. First, the strain mentioned above is the relative inter-layer strain, composed of the strain tensor of each layer composing the bilayer structure. Second, the strain tensor of layer $i$ has the form:

$$\overleftrightarrow{\epsilon_i} = \frac{1-\nu_i}{2} S_i \begin{pmatrix} 1 & 0 \\ 0 & 1 \end{pmatrix} + \frac{1+\nu_i}{2} S_i (\cos\phi_s\,\sigma_x + \sin\phi_s\,\sigma_z)$$

Where $\nu_i$ is the Poisson ratio of that layer. If we assume the two layers are strained along the same direction (both have the same $\phi_s$) we would get a relative strain tensor of a similar form:

$$\overleftrightarrow{\epsilon} = \frac{1-\nu}{2} S \begin{pmatrix} 1 & 0 \\ 0 & 1 \end{pmatrix} + \frac{1+\nu}{2} S (\cos\phi_s\,\sigma_x + \sin\phi_s\,\sigma_z)$$

For $S = S_2 - S_1, \nu = \frac{\nu_2 S_2 - \nu_1 S_1}{S_2 - S_1}$, and we can continue with the same analysis, as before with $\epsilon_s = -\nu S$, $\epsilon_c = S$. For the simplest case of a homo-bilayer structure in the low twist angle limit, such that $\delta = 0, \theta \ll \nu = \nu_1 = \nu_2$ we get a critical strain of approximately $\frac{2}{\sqrt{\nu}} \sin\frac{\theta}{2}$. This realistic critical behavior in twisted bilayer graphene is demonstrated in Fig. S7d-f and in Supplementary Video 6.

## S 9. Large rhombohedral domains as meta-stable states

Figure 3a presented a histogram of radii of curvature of TDBG rhombohedral domains for cases of controlled electrostatic environment near CNP. Throughout the fabrication process, before the stacks were contacted with metallic electrodes, and specifically for stacks that were cut with anodic oxidation (see methods section), we persistently observed isolated large rhombohedral domains with significantly larger $\kappa^{-1}$ values, up to $\kappa^{-1} = 5 \cdot 10^3\ nm$. Fig. S8a extends the histogram of Fig. 3a to include these cases as well (red bins). These latter large domains are stable at room temperature but disappear under thermal annealing, as shown in Fig. S8b-c, suggesting they are meta-stable states. In this example we observe the annihilation of large rhombohedral domains by vacuum thermal annealing (at 350℃) which is a common step in the fabrication process of vdW heterostructures. Fig. S8b presents the nearfield phase of a TDBG device before thermal annealing, showing a few large ABCA domains (dark regions marked with red arrows) and a strained structure. After annealing (Fig. S8c) the large domains vanish, as the overall moiré pattern becomes more uniform, and seems to indicate a reduced level of strain (except for the left part, where a mechanical rotator was placed, clearly deforming the pattern, as expected). This behavior, of annihilation of large rhombohedral domains by thermal annealing, seems to be a common theme in these structures. A more vivid demonstration of the meta-stability of these large domains is presented in the insets Fig. S8d-f, which are three sequential mid-IR nearfield scans of the same region (see methods). Due to optical field enhancement by the metallic tip the region under the tip experiences excessive heating, which likely drives this transition. The light-induced collapse of one of the domains (red arrows in inset Fig. S8e) suggests the scanning tip acted as a local thermal annealer attesting to the meta-stable nature of the domain. This can be seen by comparing domain structure in Fig. S8e above arrows (before collapse) to an earlier scan (Fig. S8d) and below arrows (after collapse) to a later scan (Fig. S8f).

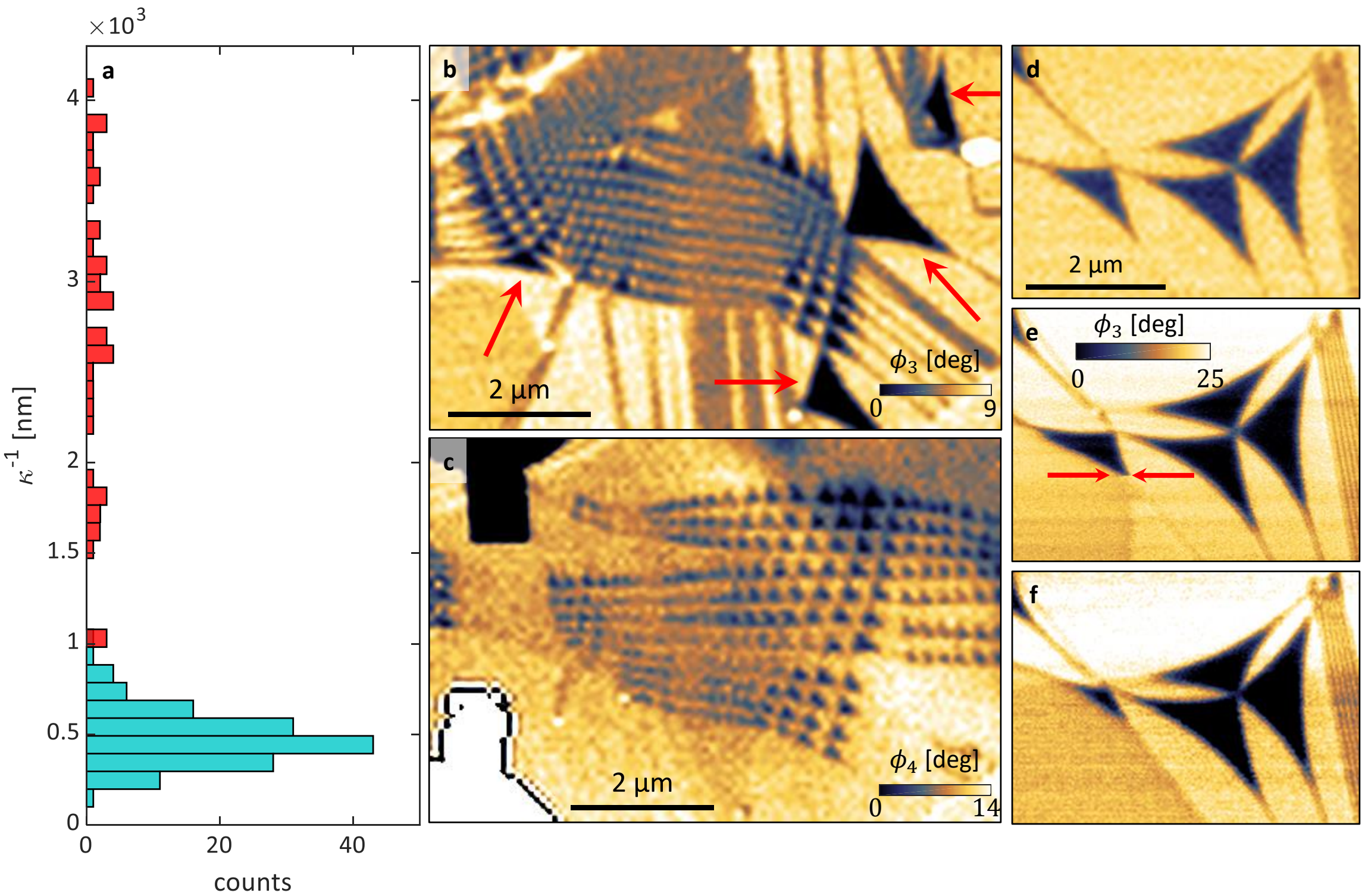

**Figure S8|Experimental indication of meta-stability of rhombohedral domains in TDBG. a,** Curvature histogram across all domains measured in this work, including those away from CNP. On top of the main cluster discussed in main-text (turquoise bins) there is a tail of large domains (red bins), that appear to be meta-stable. **b-c,** Effect of vacuum thermal annealing of a device shows a dramatic effect on rhombohedral domains pointing to their meta-stability. Before annealing (**b**) there exist large rhombohedral domains (dark) which completely vanish after annealing (**c**). The annealing also leads to strain relief which makes a more regular moiré super-lattice closer to the ground state. The added structure on the left top and bottom of (**c**) are rotators placed on top of the stack for unrelated purposes. **d-f,** 3 subsequent nearfield phase imaging scans of meta-stable ABCA domains, representing the large $\kappa^{-1}$ tail of the histogram, and demonstrating meta-stability by observing tip-induced annealing of one of the domains mid-scan**.** Red arrows (**e**) mark a sudden change in domain shape as the tip scans from top to bottom, which in later (**f**) reveals a collapsed domain. **d-f** share a color-map and scale-bar.

Attempting to pin-down the source of this meta-stability, one can come up with a few possible explanations. One obvious explanation could be that the stacks are charged, and therefore the ground-state curvature deviates from the CNP prediction presented in Fig. 3. If, for instance, during the anodic oxidation process the top graphene layer was oxidized, that would result in a charge transfer that would lead to a defect-induced doping, as we believe is the case for Fig. 3e. However, the meta-stability of the large domains, as indicated by their collapse under heating, requires a different mechanism.

Let's further explore meta-stability in this system computationally by a closer look at the relaxation process through the calculation. We track the domain walls of the system through different gradient descent steps of the simulation as it approaches toward an energy-minimizing solution. This mimics the actual relaxation process, as domain formation is driven by local forces governed by elastic and stacking energy terms. Fig. S9a plots the evolution of the rhombohedral domain through the last phases of the 2D relaxation calculation of Fig. 3d. The domain wall is tracked for a given step and assigned a color, corresponding to the instantaneous total energy at that iteration step. Initially a straight domain wall is formed (not shown for brevity). Only near convergence does the domain wall bend inward (decreasing $\kappa^{-1}$) to further decrease the energy. At some point two SDWs partially collapse to form a DDW, before converging to the ground-state. Observing how sensitive $\kappa^{-1}$ is to the energy density, one can expect that a small energy barrier could readily trap the system in a meta-stable state with, in principle, an arbitrarily large $\kappa^{-1}$ value. We can imagine the path in phase-space the system takes through the relaxation process. Fig. S9a suggests that close to the ground state the energy landscape is shallow, showing minimal changes in energy, while the curvature changes dramatically. We can further visualize this landscape by the presented soap-bubble model, as shown in Fig. S9b. Such an approach suggests, as expected, the existence of a steep direction (mostly along parameter $a$) and a shallow direction (mostly along parameter $\chi$) in the energy landscape. The radius of curvature strongly changes (indicated by colored circles) along the shallow valley from a straight domain walls on one end of the valley (for maximal $\chi$) down to small $\kappa^{-1}$ value of the true ground state at the other end of the valley, all throughout an energy drop of 30 $\mu eV/nm^2$. One can expect that if there were an energy barrier of that scale, it could readily result in a meta-stable state with large $\kappa^{-1}$ value. One possible source of such an energy barrier would be twist-angle inhomogeneity or strain. Fig. S9c compares the total energy averaged over a moiré unit-cell as a function of twist angle within the 2D soap-bubble model (and the parameters of the DFT-D2 approach listed before). One finds that for an experimentally observed twist angle variation the resulting energy landscape is expected to be sufficiently disordered to compete with relaxation processes along the shallow valley, and account for the required energy barrier to trap the system in a meta-stable state with large $\kappa^{-1}$ values. In that sense, thermal annealing has two effects: It may provide the activation energy to push the system beyond a barrier, or alternatively, it relieves strain and leaves the system more uniform, thus removing energy barrier altogether.

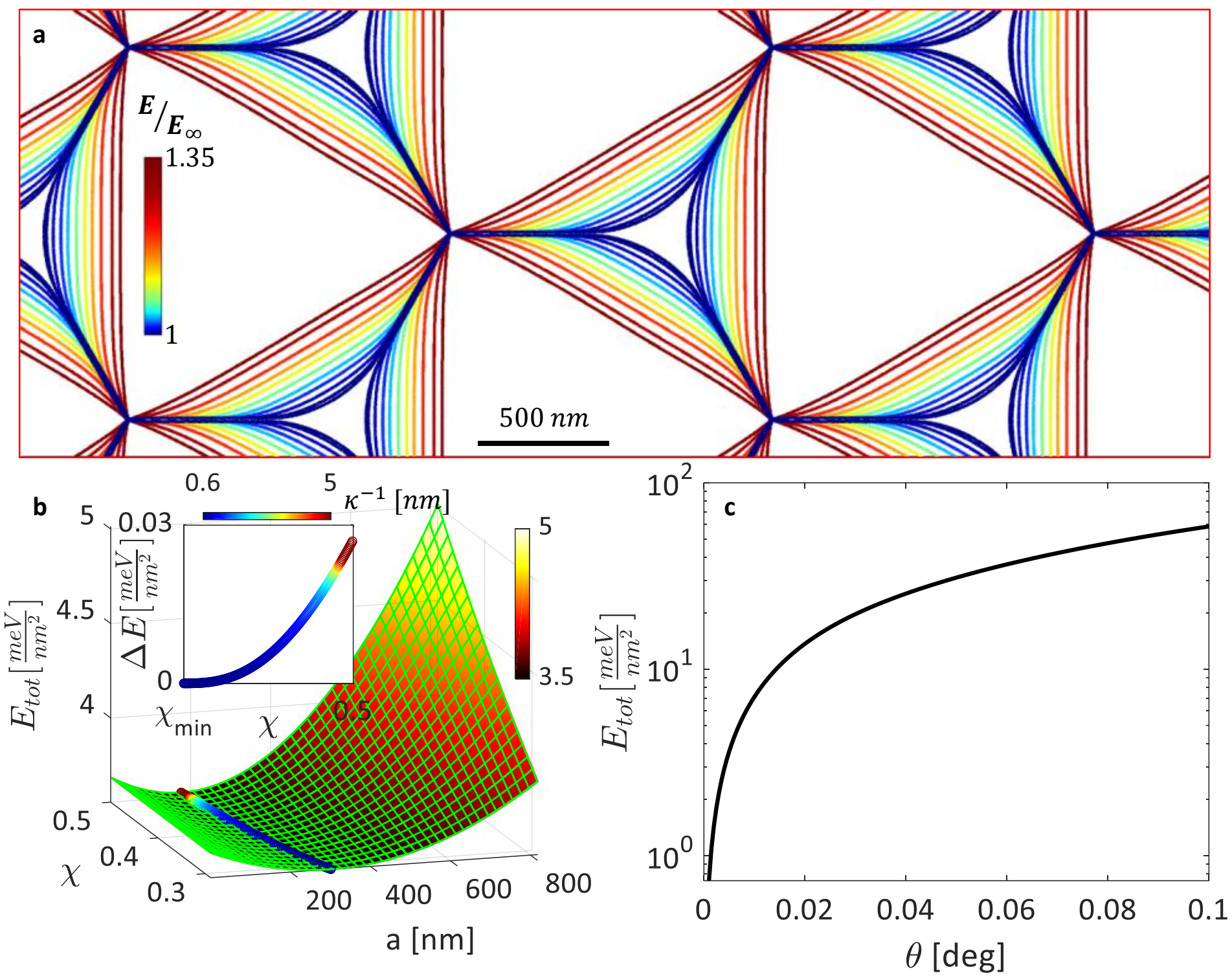

**Figure S9 | Modelling meta-stability of rhombohedral domains in TDBG. a,** Tracking domain wall shapes through the final energy minimization iterations of the calculation of Fig. 1d**.** A sharp drop in $\kappa^{-1}$ is observed as the system relaxes to its ground state. Each configuration is colored according to its total energy normalized to the ground-state energy (indicated by color-map). **b,** Total energy landscape within the 2D soap-bubble model detailed in SI section S9 for a twist angle of 0.01°, revealing a steep direction (along a) and a shallow direction (along $\chi$). Colored circles (Inset and on landscape) mark follow the shallow valley and indicate the domain radius of curvature at a given point. The parameters of DFT-D2 were used in this example (see SI section S2). **c,** Averaged (over moiré unit-cell) energy density of TDBG as a function of twist angle within the 2D soap-bubble model.